\documentclass[a4paper,UKenglish,cleveref, autoref, thm-restate,authorcolumns]{lipics-v2019}

\usepackage{latexsym,color,graphicx,amsmath,amssymb,subcaption}

\def\reals{{\mathbb R}}
\def\sph{{\mathbb S}}
\def\eps{{\varepsilon}}
\def\bd{{\partial}}
\def\A{{\cal A}}

\def\dd{{\sf dist}}
\def\ddv{{\sf d_v}}

\def\qmx{{q_{\rm max}}}

\def\nn{{\bf n}}
\def\eeta{{\bar{\eta}}}

\title{Duality-based approximation algorithms for depth queries and maximum depth}

\titlerunning{Duality-based approximation algorithms for depth queries and maximum depth}
\author{Dror Aiger}%
{Google} {aigerd@google.com}%
{}%
{}
\author{Haim Kaplan}%
{School of Computer Science, Tel Aviv University, Tel~Aviv, and Google}%
{haimk@tau.ac.il}%
{}%
{Partially supported by ISF grants 1841/14, 1595/19, by grant 1367/2016
   from the German-Israeli Science Foundation (GIF), and by Blavatnik Research Fund in Computer Science at Tel Aviv University.}%
\author{Micha Sharir}%
{School of Computer Science, Tel Aviv University, Tel~Aviv,
   Israel}%
{michas@tau.ac.il}%
{}%
{Partially supported by ISF Grant 260/18,  by grant 1367/2016
   from the German-Israeli Science Foundation (GIF), and by
   Blavatnik Research Fund in Computer Science at Tel Aviv University.}
\authorrunning{D. Aiger, H. Kaplan, M. Sharir}

\Copyright{D. Aiger, H. Kaplan, M. Sharir}


\begin{document}
\nolinenumbers
\maketitle

\begin{abstract}
We design an efficient data structure for computing a suitably defined approximate
depth of any query point in the arrangement $\A(S)$ of a collection $S$ of $n$
halfplanes or triangles in the plane or of halfspaces or simplices in higher dimensions.
We then use this structure to find a point of an approximate maximum depth in $\A(S)$.
Specifically, given an error parameter $\eps>0$, we compute, for any query
point $q$, an underestimate $d^-(q)$ of the depth of $q$,
that counts only objects containing $q$, but is allowed to exclude objects when $q$ is $\eps$-close to their boundary.
Similarly, we compute
an overestimate $d^+(q)$
that counts all objects containing $q$ but may also count objects that do not contain $q$ but $q$ is $\eps$-close to their boundary.

Our algorithms for halfplanes and halfspaces
are linear in the number of input objects and in the number of queries,
and the dependence of their running time on $\eps$ is considerably better
than that of earlier techniques.
Our improvements are particularly substantial for triangles and
in higher dimensions.

We use a primal-dual technique
similar to the algorithms for computing $\eps$-incidences in~\cite{AKS}.
Although the simplest setup of halfplanes in $\reals^2$ is not much different
from the algorithms for computing $\eps$-incidences in~\cite{AKS}, here
we apply this technique for the first time also in higher dimension. Furthermore,
the  cases of triangles in $\reals^2$ and of  simplices in
higher dimensions are considerably more involved, because the dual part of our structure requires
(for triangles and simplices) a multi-level approach, which is
problematic in our context. The reason is that in our setting progress is achieved 
by shrinking the bounding box of the subproblem (rather than the number of objects 
it contains), and this progress is lost when we switch from one dual level to the next.
Although the depth problem is, in a sense, a dual variant of the range counting problem,
these new technical challenges that we address here,
 do not have matching counterparts in the range searching context.

Our algorithms are easy to implement,
and, as we demonstrate, are fast in practice, and compete very favorably with
other existing techniques. We discuss several applications to various
problems in computer vision and related topics, which have motivated our study.
\end{abstract}

\section{Introduction} \label{sec:intro}

The \emph{depth}, $d(q)$, of a point $q$ in an arrangement of a set $S$ of $n$ simply-shaped closed
objects in $\reals^d$ is the number of objects in $S$ that contain $q$.
We consider approximate versions of the following two problems (1) Preprocess
$S$  into a data structure, such that for any query point
$q \in \reals^d$, we can efficiently report its depth in $\A(S)$.
(2) Compute a point in $\reals^d$ of maximum depth in the arrangement $\A(S)$.
We present approximate solutions to these problems, reviewed in Section~\ref{sec:contrib},
that are considerably more efficient than existing solutions, or than suitable
adaptations thereof. Both problems have many applications; we describe some of them in Section \ref{sec:applications}.

In this paper we only consider the (basic) cases where the objects in $S$ are halfspaces
or simplices. A straightforward (but typically costly) way of answering depth queries
is to construct the arrangement $\A(S)$, label each of its faces (of any dimension)
with the (fixed) number of objects that contain the face, and preprocess the arrangement
for efficient point location queries. Computing a point of exact maximum depth is then
performed by iterating over all faces of $\A(S)$, and returning (any point of) a face of maximum depth.

For halfspaces, we can dualize the problem, turning
$S$ into a collection $S^*$ of $n$ points in $\reals^d$, and a query point
$q$ into a suitably defined halfspace. We then need to preprocess $S^*$
into a data structure for halfspace range counting queries. The standard theory for
the latter problem (which is summarized, e.g., in \cite{Ag,AE}) admits
a trade-off between the storage (and preprocessing cost) of the structure
and the query time. Roughly, if one allows $s$ storage, the query cost is
close to $n/s^{1/d}$ (and the preprocessing cost is close to $s$),
so a fast query time requires storage (and preprocessing)
about $n^d$. Alternatively, if we expect to perform $m$ queries, a suitable choice
of $s$ makes the running time close to $O(m^{d/(d+1)}n^{d/(d+1)})$.

For depth with respect to simplices we can also dualize the problem.
Every simplex $\Delta$ dualizes to a tuple of points $h_1^*,\ldots, h_{d+1}^*$,
where each $h_i^*$ is dual to a hyperplane $h_i$ supporting a facet of $\Delta$.
The query $q$ translates to a hyperplane $q^*$. The depth of $q$ is equal to
the number of tuples $(h_1^*,\ldots, h_{d+1}^*)$ such that $h^*_i$ is above/below
$q^*$ if and only if $\Delta$ is below/above $h_i$, for $i=1,\ldots,d+1$.
This problem can be solved using a multi-level halfspace range counting data
structure with tradeoffs similar to those described above.

It follows that answering exact depth queries, with fast processing of a query,
seems to require preprocessing time and storage about $n^d$.
Finding a point of maximum depth also takes time close to this bound.
Moreover, the cost of answering $m$ queries on $n$ objects is superlinear,
getting close to the naive upper bound $O(mn)$ as $d$ grows.
This motivates the design of approximation schemes to tackle these problems.

%%%%%%%%%%%%%%%%%%%%%%%%%%%%%%%%%%%%%%%%%%%%%%%%%%%%%

\smallskip
\noindent
{\bf Previous work on approximate depth.}
If the class of objects has small VC dimension (as do halfspaces and simplices),
we can sample a subset $R$ of $S$ of size proportional to $1/\eps^2$, for a
prescribed error parameter $\eps>0$, apply the trivial solution described above
to $R$,  for any query point $q$, rescale the resulting depth by $|S|/|R|$, and obtain an approximation
of the true depth of $q$, within an additive error of $\pm\eps |S|$ \cite{HarPeled:book,Haussler1987}.
Unfortunately, an additive error of $\eps |S|$ does not suffice for many of the applications.

Approximation algorithms that achieve a $(1\pm \eps)$ relative error have been
studied extensively in the dual setting of halfspace range counting, and mainly
in two and three dimensions \cite{AC09,AHP,KaplanRS11}.
We recall, that in three dimensions, an exact query takes
$O(n^{2/3})$ time if we allow only linear space, and for a logarithmic query time we need cubic space.
This line of work culminated in the work of Afshani and Chan \cite{AC09},
who construct a data structure of $O(n)$ expected size in $O(n)$ expected time,
that answers an approximate depth query in $O(\log(n/k))$ expected time, where $k$ is the true depth of the query.
(Their bounds also depend polynomially on $1/\eps$ in a way which was not made explicit in \cite{AC09}.)

Still in the dual setting of approximate range counting, Arya and Mount \cite{AM00},
and later Fonseca and Mount \cite{FM10} considered a different notion of approximation,
closely related to the one that we define here (for depth). Specifically, in these works,
in the context of range counting, the query ranges are treated as ``fuzzy'' objects,
and points too close to the boundary of a query object, either inside or outside,
can be either counted or ignored. Arya and Mount \cite{AM00} gave an $O(n)$-size
data structure that can answer counting queries in convex ranges (of constant complexity)
in $O(\log n + \frac{1}{\eps^{d-1}})$ time ($\eps$ here measures % the allowed `fuzziness', namely 
the distance to the boundary within which points can be either counted or ignored).
Fonseca and Mount \cite{FM10} gave an octree-based data structure that can be constructed
in $O\left(n + \frac{(\log(1/\eps))^{d+1}}{\eps^d}\right)$ time and then can be used to
count the number of hyperplanes at approximate distance at most $\eps$ from a query point
in constant time
(this notion of `$\eps$-incidences' was also studied in a recent paper~\cite{AKS}).
Specifically, it counts all hyperplanes at distance at most $\eps$ from a query point,
and may count hyperplanes at distance up to $O(\sqrt{d}\eps)$. This data structure can
be extended to simplices and other algebraic surfaces, but with a higher cost (see \cite{AKKSZ}),
and also for approximate depth queries rather than approximate incidence queries.

%%%%%%%%%%%%%%%%%%%%%%%%%%%%%%
\subsection{Our contributions} \label{sec:contrib}
%\subparagraph*{Depth queries and maximum depth.}
%In this paper we present approximate algorithms for the depth problems
%(both the preprocessing-and-query and the maximum depth versions), using
%a different definition of approximation, for halfplanes and triangles
%(or polygons with a constant number of edges) in the plane, and for
%halfspaces and simplices in higher dimensions. Our technique extends
%the machinery developed in our previous work \cite{AKS}. Here is a brief overview
%of the concrete scheme that we study.
%For simplicity, we discuss here only the case of halfplanes in $\reals^2$.
%Let $S$ be a set of $n$ halfplanes in $\reals^2$. For simplicity,
%and without loss of generality, assume that all the potential queries lie
%in the unit square $Q:=[0,1]^2$. We denote by $\A(S)$ the arrangement formed by the
%bounding lines of the halfplanes in $S$. The \emph{depth} of a point $q\in\reals^2$,
%denoted by $d(q)$ (we suppress the dependence on $S$, which is generally clear
%from the context), is the number of objects of $S$ that contain $q$.
We define the following rigorous notion of approximate depth (along the lines
of the notion of approximate counting of \cite{AM00,FM10} described above).

\begin{figure}[htb]
\begin{center}
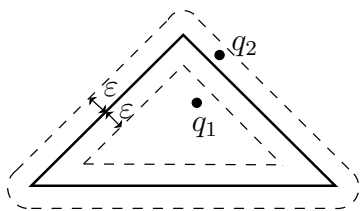
\caption{\sf The triangle is counted in both the inner $\eps$-depth and the
outer $\eps$-depth of $q_1$, but only in the outer $\eps$-depth of $q_2$.}
\label{fig:epsdepth}
\end{center}
\end{figure}

For an error parameter $\eps>0$, we define the \emph{inner $\eps$-depth} of $q$,
denoted by $d_\eps^-(q)$, to be the number of objects $s\in S$ such that
$s$ contains $q$ and $q$ lies at distance at least $\eps$ from $\bd s$, and the
\emph{outer $\eps$-depth} of $q$, denoted by $d_\eps^+(q)$, to be the number of objects
$s\in S$ such that either $s$ contains $q$ or $q$ lies (outside $s$ but) at distance
at most $\eps$ from $\bd s$. See Figure \ref{fig:epsdepth}.
%These concepts have already been considered by Arya and Mount~\cite{AM00}
%(using a somewhat different notation), albeit for the dual
%\emph{range searching} version of the problem, already mentioned above,
%where the data consists of $n$ points in $\reals^d$ and each query specifies some simple
%object, say a halfspace, and asks for the number of input points in the object.

\vspace*{-0.4cm}

We adhere to the definition of $d^+_\eps(q)$ given above, for halfplanes and halfspaces,
but modify it slightly for triangles and simplices, for technical reasons.\footnote{%
  Our techniques are not immediately suitable to work with curved objects,
  which arise around the corners of the triangles (see Figure~\ref{fig:epsdepth})
  and around the lower-dimensional faces of the simplices.}
Concretely, for a given $\eps>0$, we define the \emph{$\eps$-offset} of a simplex $\sigma$
to be the simplex $\sigma^+_\eps$ formed by the intersection of the halfspaces that contain
$\sigma$ and are bounded by the hyperplanes obtained by shifting the supporting hyperplane
of each facet of $\sigma$ by $\eps$ away from $\sigma$; see Figure~\ref{offset} for an illustration.
Then $d^+_\eps(q)$ counts the number of simplices whose $\eps$-offset contains $q$.\footnote{%
  In practice, we estimate a smaller quantity; see later in the paper.}
See Sections \ref{sec:triangles}, \ref{sec:simplices}.

We assume that all query points are restricted to the unit cube $Q=[0,1]^d$.
We are interested in a data structure that can compute efficiently, for any
query point $q$ in $Q$, a pair of integers $d^-(q)\le d^+(q)$, such that
\begin{equation} \label{dplmi}
d_\eps^-(q)\le d^-(q) \le d(q) \le d^+(q) \le d_\eps^+(q) .
\end{equation}
In a stronger form (followed in this paper), we require that (i) every object
in the inner $\eps$-depth of $q$ be counted in $d^-(q)$, (ii) every object counted
in $d^-(q)$ contain $q$, (iii) every object containing $q$ be counted in $d^+(q)$,
and (iv) every object counted in $d^+(q)$ be in the outer $\eps$-depth of $q$.
These conditions trivially imply (\ref{dplmi}).
We view $d^-(q)$, $d^+(q)$ as an underestimate and an overestimate of $d(q)$.

%In this paradigm, if a point $q$ is $\eps$-close to
%the boundary of some object $s\in S$ then, if $q$ is contained in $s$, we might fail
%to count this (shallow) containment in $d^-(q)$, and if $q$ is outside $s$, we might
%count this (false) containment in $d^+(q)$. Still, we guarantee that any object
%counted in $d^-(q)$ contains $q$, and any object that contains $q$ is counted
%in $d^+(q)$. It follows that, in the worst case,
%$d^-(q)$ might be smaller than $d(q)$ by the number of shallow containments of $q$,
%so the maximum value of $d^-(q)$ might be considerably smaller than the maximum depth
%in $\A(S)$, but the difference is at most the number of shallow containments at the
%point (or points) of maximum depth. Similarly, $d^+(q)$ might be larger than $d(q)$
%by at most the number of false containments at $q$, and the maximum value of $d^+(q)$
%might be larger than the true maximum depth by at most the number of false containments
%at the points $q$ that attain $d^+(q)$.

In this paper we give data structures for depth queries in arrangements of
halfspaces and simplices in $\reals^d$. We first focus on halfplanes and triangles
in $\reals^2$ and then extend our algorithms to higher dimensions. In handling the
cases of triangles and of higher dimensions, we need to apply a battery of additional
(novel) techniques; these techniques are easy to define and to implement, but their
analysis is involved and nontrivial. We present
data structures for approximte depth queries and then show how to use them to compute
point(s) of approximate maximum depth. The dependency of our bounds on $\eps$ is much
better than what is currently known, or than what can be adapted from known techniques
(which mostly cater to range counting queries rather than to depth queries).
Specifically, we present the following results.

%%%%%%%%%%%%%%%%%%%%%%
\begin{theorem}[Halfplanes and triangles in $\reals^2$] \label{main:query}
Given a set $S$ of $n$ halfplanes or triangles that meet the unit square, an error parameter $\eps>0$,
and the number $m$ of queries that we need (or expect) to answer, we can preprocess $S$ into
a data structure, so that its storage,
preprocessing cost, and the time to answer $m$ depth queries, are all
$
\tilde{O}\left( \frac{\sqrt{mn}}{\sqrt{\eps}} + m + n \right) ,
$
% \[
% O\left(
% \frac{\sqrt{mn}}{\sqrt{\eps}}\log^{1/2} \frac{n}{\eps m} + \left( m + n \right)
% \log\frac{1}{\eps}\right) ,
% \]
where $\tilde{O}(\cdot)$ hides logarithmic factors, and where,
for each query point $q$, we return two numbers $d^-(q)$ and $d^+(q)$ that satisfy
(\ref{dplmi}), in the stronger sense of containment discussed there.
% The bound for triangles is larger by a logarithmic factor.
\end{theorem}
%%%%%%%%%%%%%%%%%%%%%%

% In higher dimensions we get the following bounds.

%%%%%%%%%%%%%%%%%%%%%%%%%%%%%%%%%%%%%%%%%%
\begin{theorem}[Halfspaces and simplices in $\reals^d$] \label{main:query:d}
Given a set $S$ of $n$ halfspaces that meet the unit cube, an error parameter $\eps>0$,
and the number $m$ of queries that we need (or expect) to answer, we can preprocess $S$ into
a data structure, so that its storage,
preprocessing cost, and the time to answer $m$ depth queries, are all
$
\tilde{O}\left( \frac{\sqrt{mn}}{\eps^{(d-1)/2}} + m + n \right) ,
$
% \[
% O\left( \frac{\sqrt{mn}}{\eps^{(d-1)/2}} \log^{1/2}\frac{n}{m\eps^{d-1}}
% + (m + n) \log\frac{1}{\eps} \right) ,
% \]
where, for each query point $q$, we return two numbers $d^-(q)$ and $d^+(q)$ that satisfy
(\ref{dplmi}), in the stronger sense of containment discussed there. The bound for simplices is
$
\tilde{O}\left( \frac{m^{2/(d+2)}n^{d/(d+2)}}{\eps^{d(d-1)/(d+2)}} + m + n \right) .
$
% \[
% O\left( \left( \frac{m^{2/(d+2)}n^{d/(d+2)}}{\eps^{d(d-1)/(d+2)}}
% \log^{\frac{d(d-1)}{d+2}} \left( \frac{n}{m\eps^{d-1}}\right)
% + m \log\frac{1}{\eps} + n \log^{d-1}\frac{1}{\eps} \right)\log n \right) .
% \]
\end{theorem}
%%%%%%%%%%%%%%%%%%%%%%

The results for halfplanes and halfspaces are given in Sections \ref{sec:halfplanes} and \ref{sec:halfspaces}, respectively,
and the results for triangles and simplices are given in Section \ref{sec:triangles} and \ref{sec:simplices}, respectively.

All our bounds are $\tilde{O}(m+n)$, and that their dependency on $\eps$ is much better than of any
existing algorithm. Specifically, our dependence is $1/\sqrt{\eps}$ instead of $1/\eps$ in $\reals^2$,
and $1/\eps^{{(d-1)}/{2}}$ instead of $1/\eps^{d-1}$ for hyperplanes in $\reals^d$.
For depth in simplices, no explicit result was stated in the earlier works,
and our bounds are considerably better than what one could get using previous techniques.
% We do have a sublinear depedency on $n$ and $m$ that \cite{FM10} did not have.
% \micha{Remove or explain the last sentence.}

\subparagraph*{Approximate maximum depth.}
% \micha{State the theorems here.}
% \haim{I do not think we want to do this}

Our data structure can be applied to find points $q^-$ and $q^+$ in the unit cube $Q$ such that
$d^-(q^-)$ is close to $\max_{q\in Q} d^-(q)$, as well as a point $q^+$ such that
$d^+(q^+)$ is close to $\max_{q\in Q} d^+(q)$. These points depend,
among other things, on the specific way in which we define (and compute) $d^-(q)$ and $d^+(q)$,
and are not necessarily the same. Nevertheless, the deviations of $d^-(q^-)$ and $d^+(q^+)$ from
the true depths $d(q^-)$ and $d(q^+)$ (and also from the true maximum depth), are only due
to objects such that $q^-$ and $q^+$ are close to their boundaries, respectively.

To compute an approximate maximum depth, in this sense, we query our data structure with
the cell centers of a sufficiently dense grid (of cells with side length proportional to $\eps$),
and return a center of a grid cell with maximum $d^-$-value, and a (possibly different)
center with maximum $d^+$-value.
We prove that these points yield good approximations to the maximum depth, in the 
fuzzy sense used here (see, e.g., Theorem \ref{th:maxdepth-main}).
This calls for answering $m=\Theta\left(\frac{1}{\eps^d}\right)$ queries, so the
running time of this method degrades that much with the dimension, but so do
(suitable adaptations of) the earlier techniques of \cite{AM00,FM10}.
Our maximum depth algorithms for halfplanes and halfspaces are somewhat simpler, 
have smaller hidden constant factors and have smaller polylog factors in $1/\eps$.

For the cases of triangles and simplices, the dependence on $\eps$ is significantly smaller
in our algorithms. For example, in the context of (the `dual') range searching, the dependence
on $\eps$ in the algorithm of \cite{FM10} is $O(1/\eps^{2d-2})$, and in the context of depth
queries (which is not explicitly covered in \cite{FM10}), the best dependence that seems
to be obtainable from their technique is $O(1/\eps^{d(d+1)})$. In contract, the dependency
on $\eps$ in our bound, given
in Theorem~\ref{main:query} (with $m=O(1/\eps^d)$ queries) is only $O(1/\eps^d)$ (the leading
term has a slightly better dependency). We leave open the question of whether one can make do
by asking muchy fewer explicit queries in order to approximate the maximum depth.

% \micha{This is the selling para. I put it in Appendix A in full, and trimmed it here.
% It has to be merged better with the rest of the text.}
Although the simplest setup of halfplanes in $\reals^2$ is treated here in a manner
that is not too different from the algorithms for computing $\eps$-incidences in~\cite{AKS},
the other cases, of triangles in $\reals^2$ and of halfsplaces and simplices in
higher dimensions, are considerably more involved:
(i) They need a battery of additional ideas for handling multi-level structures, 
of the sort needed here, in higher dimensions.
(ii) They yield substantially improved solutions (when $n$
is reasonably large in terms of $\eps$ or when the number of queries is not too excessive).
(iii) They are in fact novel, as the depth problem, under the fuzzy model assumed here
(and in \cite{AM00,FM10}), does not seem to have been considered in the previous works.
Although the depth problem is, in a sense, a dual variant of the range counting problem,
it raises, under the paradigm followed in this paper,
new technical challenges, which do not have matching counterparts in the range
searching context, and addressing these challenges is far from trivial, as we
demonstrate in this work.

%%%%%%%%%%%%%%%%%%%%%%%%%%%%%%%%%%%%%%%%%
\subparagraph*{An overview of the technique.}
We use a primal-dual approach, similar, at high level, to the one that was used in recent
works~\cite{AKKSZ,AKS} for computing approximate incidences. To apply this technique
for approximate depth in the fuzzy model considered here, we use oct-trees both in
the primal and dual spaces, as well as an additional level of a segment tree structure for
triangles and simplices. Handling the cases of triangles and simplices requires
new ideas for combining this primal-dual approach with a multi-level data structure:
In traditional primal-dual multi-level data structures for range searching, we reduce
the size of the problem at each recursive step, and progress is measured by the number of points
and objects that each step involves. Here, in contrast, progress is made by reducing
the box size in which the subproblem ``lives''. This approach is problematic when
the structure consists of several levels, as the features stored in one level are
different from those stored at previous levels, and are not necessarily confined to the
same smaller-size region that contains the previous features. A novel feature that we need to address
is to ensure that this gain is not lost when we switch to the dual space, or move to 
a different level of the structure in the dual case.

% \haim{Here is an attempt to elaborate some more on this duality issue..}
Specifically, in the dual setting for depth in an arrangement of simplices,
checking whether a query point $q$ is contained in a simplex $\sigma$
amounts to checking whether the dual halfspace $q^*$ is on the correct side of
each point in the tuple $(h^*_1,\ldots,h^*_d)$ of points dual to the supporting
hyperplanes of the facets of $\sigma$. The technical challenge here is that
we need to test this property for each index $j=1,\ldots,d$ separately, meaning that,
for each $j$, the $j$-th dual level needs to handle the points $h^*_j$, over
all simplices, and none of these points need to bear any tangible relationship to
the preceding points $h^*_1,\ldots,h^*_{j-1}$ of the same simplex. This means
that the proximity gain that we get by reducing the size of the box of, say, the first
dual level, that contains the first dual points $h^*_1$, is lost when turning to preprocess the second
dual points $h^*_2$, and this continues through all dual levels.
% Once we leave the first level we know that all points $p_1^*$ in a box $B$
%of certain size are on the right side of $q^*$. However, when we switch to the
%next level the corresponding points $p_2^*$ are not guaranteed to be in $B$
%and our proximity gain is lost.
We overcome this problem in the plane for
triangles by avoiding a dual multilevel structure altogether. But in
higher dimensions all we can do is reduce the number of levels, but not
avoid them completely, and this is the reason for our strange-looking bounds
for simplices in Theorem \ref{main:query:d}. Our work leaves open the challenge
of improving these bounds (possibly even getting the same bounds as we have for halfspaces).

Here is a brief overview of our approach (described for halfplanes in $\reals^2$,
for simplicity). We construct in the primal plane (over the unit square $Q$) a coarse
quadtree $T$, up to subsquares of size $\delta_1$, for a suitable
parameter $\eps\le\delta_1\le 1$. We pass the lines bounding the given halfplanes
through $T$, and store with each square $\tau$ of $T$ the number of halfplanes
that fully contain $\tau$ but do not fully contain the parent square of $\tau$.
Squares that are not crossed by any bounding line become `shallow' leaves and are not
preprocessed further. For each bottom-level leaf $\tau$, we take the set of
halfplanes whose bounding line crosses $\tau$, dualize its elements into points,
and process them in dual space, using another quadtree, expanded until we reach
an accuracy (grid cell size) of $\eps$. (See below for the somewhat subtle details of the dual quadtrees.)

We answer a query with a point $q$ by searching with $q$ in the primal quadtree,
and then by searching with its dual line $q^*$ in the corresponding secondary tree.
The values $d^-(q)$ and $d^+(q)$ that we return are the sum of various counters
(such as those mentioned in the preceding paragraph)
stored at the nodes of both primal and dual trees that the query accesses,
with more counters added to $d^+(q)$ than to $d^-(q)$.

Handling triangles is done similarly, except that we first replace them by right-angle
axis-aligned vertical trapezoids whose lower sides are horizontal and lie all
on a common horizontal line (see below, and refer to Figure~\ref{tri:right}).
Each triangle is the suitably defined `signed union' (involving unions and differences)
of its trapezoids.
We construct a segment tree over the $x$-spans of the trapezoids, which allows us
to reduce the problem to one involving `signed' halfplanes (see later),
which we handle similarly to the way described above.
This bypasses the issue of having to deal with round corners of the region 
at distance at most $\eps$ from a triangle, but it comes at the cost of
potentially increasing the number of triangles that will be counted in
$d^+(q)$. We control this increases using additional insights into the
structure of the problem.
%The segment tree makes the performance of the algorithm slightly less efficient,
%increasing the bounds by a logarithmic factor.
%\micha{Do we want to comment on the `bug' and its solution here? Probably not?}

The extensions to higher dimensions are conceptually straightforward, but the
adjustment of the various parameters, and the corresponding analysis of the
performance bounds, are far from simple. The resulting bounds (naturally) become worse
as the dimension increases. Nevertheless, they are still only $\tilde{O}(m+n)$,
and are much faster, in their dependence on $\eps$, than the simpler
solution that only works in the primal space (as in, e.g., \cite{AM00}, or as
can be derived from the analysis in \cite{FM10}).

Due to lack of space we postpone many details of our structures and analysis to the appendices.

\subsection{Applications and implementation} \label{sec:appl}

Finding the (approximate) maximum depth is a problem that received attention
in the past. See Aronov and Har-Peled~\cite{AHP}, Chan~\cite{chan1996} and references 
therein for studies of this problem (under the model of an $\eps$-relative approximation
of the real depth) and of related applications. 

In many \emph{pattern matching} applications, we seek
some transformation that brings one set of points (a pattern) as close as possible to
a corresponding subset of points from a model. Each possible match between points $a$,
$b$, up to some error, generates a region $R_{a,b}$ of transformations that 
bring $a$ close to $b$. Finding a tranformation with maximum depth among these regions
gives us a transformation with the maximum number of matches. The dimension of the 
parameter space of transformations is typically low (between $2$ and $6$).

%The two kinds of approximations, the depth approximation in~\cite{AHP}\cite{chan1996} and the ``fuzzy'' objects used in this paper, are both
%valid for most of the applications. Often, a solution which is spatially close to the optimum, is better than one that is close to the maximum
%depth but can be arbitrary far away from the optimal solution.

% In computer vision, several optimization problems can be formulated as a problem of finding the maximum depth in (the arrangement of) a set of
% objects in some parameter space, and the deepest point is commonly found by subdivision in transformation space (quadtree-like). In these
% applications, the dimension of the parameter space is low (say $2$-$6$). An important class of related problems that have been addressed using
% branch-and-bound techniques is 

%In the general formulation, we are given two point sets $A$ and $B$ in $\reals^3$ or $\reals^2$ and some $\delta$, and the goal is to find a linear transformation of $A$ that maximizes the number of points in $A$ which are transformed to a distance at most $\delta$ (in $L_\infty$) from a point in $B$. This can be solved (for simplicity, assuming that the minimum pairwise distance in $A$ is larger than $\delta$) by computing the deepest point in an arrangement of convex polytopes in the transformation space, each polytope is the intersection of constant number of linear constraints. The subdivision applied by previous methods is equivalent to the simple quadtree we described above. We give more efficient solutions.

Geometric matching problems of this kind are abundant in computer vision and related 
applications; see~\cite{Breuel2003,OKO2009} and references therein. 
For example, many \emph{camera posing} problems can be formulated as a maximum incidences problem~\cite{AKKSZ},
or a maximum depth problem. In~\cite{FLO2015}, the problem of finding the best translation between two cameras 
is reduced to that of finding a maximum depth among triangles on the unit
sphere (that can be approximated by triangles in $\reals^2$). The optimal relative pose problem
with unknown correspondences, as discussed in~\cite{FLO2016}, is solved by reducing it
to the same triangle maximum depth problem on a sphere.

In another set of applications, using maximum depth as a tool, one can solve 
several \emph{shape fitting} problems with outliers, as studied in Har-Peled and Wang~\cite{HW04}. 

Answering depth queries, rather than seeking the maximum depth, is also common in these areas.
% The problems need to ask for depth queries arise when one needs to evaluate a (sparse) set of given hypotheses. The depth in the arrangement of
% a set of objects in some transformation space is the quality of the hypothesis. 
In many computer vision applications, if the fraction of inliers is reasonably large, 
a classical technique that is commonly used is RANSAC~\cite{RANSAC}, which generates a 
reasonably small set of candidate transformations, by a suitable procedure that samples from the input,
and then tests the quality of each candidate against the entire data, where each such test amounts
to a depth query.

Finally, our technique is fairly easy to implement, very much so when compared 
with techniques for exact depth computation. We report (in the appendix) on an implementation
of our technique for the case of halfplanes in $\reals^2$, and on its efficient performance in practice.

\label{sec:applications}

%%%%%%%%%%%%%%%%%%%%%%%%%%%%%%%%%%%%
\section{Approximate depth for halfplanes} \label{sec:apxhalf}
\label{sec:halfplanes}

To illustrate our approach, we begin with the simple case where $S$ is a collection of
$n$ halfplanes in $\reals^2$. We construct a data structure that computes numbers
$d^-(q)$, $d^+(q)$ that satisfy (\ref{dplmi}), for queries $q$ in the square
$Q=[0,1]^2$, and for some prespecified error parameter $\eps>0$.
We denote by $\ell_h$ the boundary line of a halfplane $h\in S$.

In Appendix~\ref{app:apxhalf} we first present a `naive' approach for handling this
problem. It requires $O\left(\frac{n}{\eps}\right)$ preprocessing and answers a query
in $O\left(\log\frac{1}{\eps}\right)$ time. Here we present a faster construction
(in terms of its dependence on $\eps$) that uses duality. We use standard duality
that maps each point $p=(\xi,\eta)$ to the line $p^*:\; y=\xi x-\eta$, and each line
$\ell:\;y=cx+d$ to the point $\ell^*=(c,-d)$. This duality preserves the vertical
distance $\ddv$ between the point and the line; that is, $\ddv(p,\ell) = \ddv(\ell^*,p^*)$.
We want the vertical distance to be a good approximation of the actual distance. While not true
in general, we ensure this by partitioning the set of boundary lines into
$O(1)$ subsets, each with a small range of slopes, and by repeating the algorithm
for each subset separately.

We construct a standard primal (uncompressed) quadtree $T$ within
$Q$. For $i\ge 0$, let $T^i$ denote the $i$-th level of $T$.
Thus $T^0$ consists of $Q$ as a single square, 
% $T^1$ consists of four subsquares of side length $1/2$, 
and in general $T^i$ consists of $4^i$ subsquares of side length $1/2^i$.
For technical reasons, it is advantageous to have the squares at each level pairwise
disjoint, and we ensure this by making them half-open.
We construct the tree up to level $k = \log \frac{1}{\delta_1}$, for some parameter
$\eps\le \delta_1 \le 1$, so each leaf $v$ in $T^k$ represents a square $\tau_v$
of side length $\delta_1$. For each node $v$ of $T$ (other than the root), we
maintain a counter $c(v)$ of the number of halfplanes $h$ that fully contain $\tau_v$
but $\ell_h$ crosses the parent square $\tau_{p(v)}$ of $\tau_v$.

\begin{figure}[htb]
\begin{center}
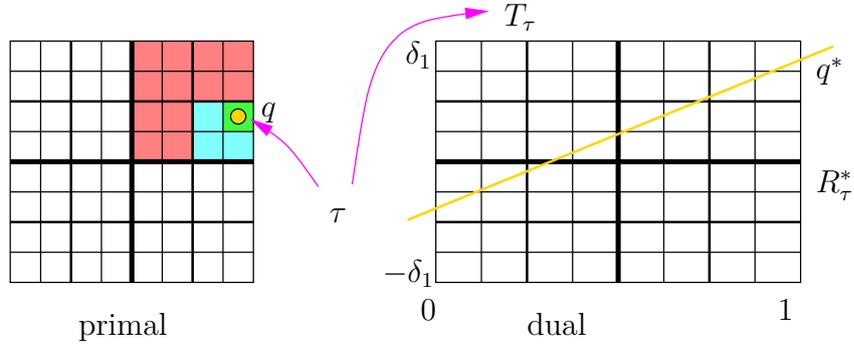
\caption{\sf A schematic illustration of the structure.}
\label{sketch}
\end{center}
\end{figure}

For each deep leaf $v\in T^k$,
we pass to the dual plane and construct there a dual quadtree on the set of
points dual to the boundary lines that cross $\tau_v$.
(Only leaves at the bottom level require this dual construction.)
See Figure~\ref{sketch} for a schematic illustration.

Let $\tau=\tau_v$ be a square associated with some bottom-level leaf $v$ of $T^k$.
Let $S_\tau \subseteq S$ be the subset of halfplanes $h$ whose boundary line
$\ell_h$ crosses $\tau$. We partition $S_\tau$ into four subsets
according to the slope of the boundary lines of the hyperplanes. 
Each family, after an appropriate
rotation, consists only of halfplanes whose boundary lines have slopes in $[0,1]$.
We focus on the subset where the original boundary lines have slope in $[0,1]$,
and denote it as $S_\tau$ for simplicity. The treatment
of the other subsets is analogous. The input to the corresponding dual problem at $\tau$
is the set $S_\tau^*$ of points dual to the boundary lines of the halfplanes in $S_\tau$.
In general, each $\tau$ has four dual subproblems associated with it.

We assume without loss of generality that $\tau = [0,\delta_1]^2$.
It follows from our slope condition that the boundary lines of the halfplanes in
$S_\tau$ intersect the $y$-axis in the interval $[-\delta_1,\delta_1]$.
Therefore, by the definition of the duality transformation, each dual point
$h^* \in S_\tau^*$ lies in the rectangle $R_\tau^* = [0,1]\times [-\delta_1,\delta_1]$.
Any square other than $\tau$ is treated analogously, except that the duality has to
be adjusted by a suitable shift.
% (see the appendix for the easy details).

We store the points of $S_\tau^*$ in a dual pruned quadtree $T_\tau$, whose root
corresponds to $R_\tau^*$, and for each $i$, its $i$-th level $T^i_\tau$
corresponds to a partition of $R_\tau^*$ into $2^i \times 2^i$ congruent rectangles,
each of side lengths $(1/2^i) \times (2\delta_1/2^i)$. We stop the construction
when we reach level $k^* = \log \frac{4}{\delta_2}$, for $\delta_2 = \eps/\delta_1$.
At this level, each rectangle associated with a leaf $u$ is of width $\delta_2/4$
and of height $\delta_1 \delta_2/2 = \eps/2$.

Consider a query point $q\in \tau$ and let $q^*$ be its dual line. Let $h$ be a
halfplane in $S_\tau$ and let $h^*$ be its dual point (that is, the point dual to
its boundary line). Now $q$ lies in $h$ if and only if $h^*$ lies in an appropriate
side of $q^*$: this is the upper (resp., lower) side if $h$ is an upper (resp., lower)
halfplane. We therefore encode the direction (upper/lower) of $h$ with $h^*$, by
defining $h^*$ to be \emph{positive} if $h$ is an upper halfplane and \emph{negative}
if $h$ is a lower halfplane. Each node $u$ of $T_\tau$ stores two counters $c^+(u)$
and $c^-(u)$ of the positive and negative points, respectively, of $S_\tau^*$ that
are contained in the rectangle represented by $u$.

To answer a query with a point $q$ (consult Figure~\ref{sketch}), we first search the primal quadtree
$T$ for the leaf $v$ such that $q\in \tau_v$. If $v$ is a shallow leaf,
we stop the process and output the sum of the counters $c(u)$ over all nodes $u$
on the search path to $v$, inclusive; in this case we obtain the real depth of $q$.
Otherwise, we search in the dual quadtree $T_{\tau_v}$ with the line $q^*$, and
sum the counts $c^+(u)$ of all nodes $u$ whose rectangle lies above $q^*$ but
the rectangle of the parent of $u$ is crossed by $q^*$, and the counts $c^-(u)$
of all nodes $u$ whose rectangle lies below $q^*$ but the rectangle of the parent
of $u$ is crossed by $q^*$. We denote by $C^-(q)$ and $C^+(q)$ these two respective
sums. Let $C(v)$ be the sum of the counters $c(u)$ in the primary tree of all nodes
$u$ along the path from the root to $v$. We set $d^-(q) := C(v) + C^-(q) + C^+(q)$,
and set $d^+(q)$ to be $d^-(q)$ plus the sum of all the counters $c^+(u)+c^-(u)$
of the leaves $u$ of $T_{\tau_v}$ that $q^*$ crosses.

\subparagraph*{Correctness.}
The correctness of this procedure (i.e., establishing (\ref{dplmi}) is argued as follows.
%; see Appendix~\ref{app:apxhalf} for a proof.
%%%%%%%%%%%%%%%%%%%%%%%%%%%%%%%%
\begin{lemma} \label{hs-correct}
(a) For any query point $q$ we have $d^-(q) \le d(q) \le d^+(q)$. \\
(b) Let $h \in S$.
If $q$ lies in $h$ at distance $\ge\eps$ from $\ell_h$ then $h$ is counted in $d^-(q)$. \\
(c) If $h$ is counted in $d^+(q)$ then the distance between $q$ and $h$ is at most $\eps$.
\end{lemma}
%%%%%%%%%%%%%%%%%%%%%%%%%%%%%%%%

\subparagraph*{Preprocessing and storage.}
A straightforward analysis shows that the total construction time
and storage are dominated by the cost of constructing the dual quadtrees,
which is $O\left(\frac{n}{\delta_1} \log \frac{1}{\delta_2}\right)$.

When we answer a query $q$, it takes $O\left(\log \frac{1}{\delta_1}\right)$ time to
find the leaf $v$ in $T$ whose square $\tau_v$ contains $q$, and then,
assuming $v$ to be a bottom-level leaf,
$O\left(\frac{1}{\delta_2}\right)$ time to trace $q^*$ in
$T_{\tau_v}$ and add up the appropriate counters. The total cost of a query is thus
$O\left( \frac{1}{\delta_2} + \log \frac{1}{\delta_1} \right)$,
and the total time for $m$ queries is
$O\left(m \left(\frac{1}{\delta_2} + \log \frac{1}{\delta_1}\right)\right)$.
It is easy to see that the term $\log\frac{1}{\delta_1}$ dominates only when $\delta_2$
is very close to $1$. Specifically this happens when
$\frac{1}{\log\frac{1}{\eps}} \le \delta_2 \le 1$.

\subparagraph*{Answering $m$ queries.}
Let $m$ denote the number of queries that we want (or expect) to handle.
The values of $\delta_1$ and $\delta_2$ that nearly balance the construction time with the
total time for $m$ queries, under the constraint that $\delta_1\delta_2=\eps$, are
(ignoring the issue of possible dominance of the term $\log\frac{1}{\delta_1}$
in the query cost)
$\delta_1 = \tilde{O}\left(\sqrt{\frac{n\eps}{m}} \right)$ and
$\delta_2 = \tilde{O}\left(\sqrt{\frac{m\eps}{n}} \right)$,
% $\delta_1 = \sqrt{\frac{n\eps}{m}} \sqrt{\log \frac{n}{\eps m}}$ and
% $\delta_2 = \sqrt{\frac{m\eps}{n}}\frac{1}{\sqrt{\log \frac{n}{\eps m}}}$,
and the cost is then
${\displaystyle \tilde{O}\left( \frac{\sqrt{mn}}{\sqrt{\eps}} \right)}$.
% ${\displaystyle O\left( \frac{\sqrt{mn}}{\sqrt{\eps}} \log^{1/2}\frac{n}{\eps m} \right)}$.
For this to make sense, we require $\eps \le \delta_1,\;\delta_2 \le 1$, meaning that
$n\eps \le m \le \frac{n}{\eps}$.
% ${\displaystyle n\eps \log\frac{n}{\eps m} \le m \le \frac{n}{\eps} \log\frac{n}{\eps m}}$.
The situations where $m$ falls out of this range are easy to handle,
%(see the appendix),
 and yield the additional terms
$\tilde{O}(n+m)$,
% $O\left(n\log\frac{1}{\eps}\right)$ and $O\left(m\log\frac{1}{\eps}\right)$,
for the overall bound
$
\tilde{O}\left( \frac{\sqrt{mn}}{\sqrt{\eps}} + n + m \right) .
$
% \begin{equation} \label{eq:bd}
% O\left( \frac{\sqrt{mn}}{\sqrt{\eps}} \log^{1/2}\frac{n}{\eps m}
% + n \log\frac{1}{\eps} + m \log\frac{1}{\eps}\right) .
% \end{equation}
This completes the proof of Theorem \ref{main:query} for halfplanes.

%%%%%%%%%%%%%%%%%%%%%%%%%%%%%%%%%%%%%%%%%%%%%%
\subparagraph*{Approximating the maximum depth.}
We can use this data structure to approximate the maximum depth as follows.
For each primal $\frac{\eps}{2\sqrt{2}} \times \frac{\eps}{2\sqrt{2}}$
grid square $\sigma$, pick its center $q_\sigma$, compute
$d^-(q_\sigma)$ and $d^+(q_\sigma)$, using our structure,
and report the square centers $q^-$ and $q^+$ attaining  $d^-=\max_\sigma d^-(q_\sigma)$ and $d^+=\max_\sigma d^+(q_\sigma)$.
The number of queries is $m = O\left( 1/\eps^2 \right)$. The following theorem specifies
lower bounds on the depths of these centers. Note that  the lower bound provided for
$d^+(q^+)$ is larger but $d^+(q^+)$ counts also ``close'' false containments. 
Whether this is desirable may be application dependent.
%%%%%%%%%%%%%%%%%%%%%%%%%
\begin{theorem} \label{th:maxdepth-main}
Let $S$ be a set of $n$ halfplanes in $\reals^2$ and let $\eps>0$ be an error parameter.
We can compute points $q^-$ and $q^+$ in $Q=[0,1]^2$, such that $d^-(q^-)$ and $d^+(q^+)$ closely approximate
the maximum depth in $\A(S)$ (within $Q$), in the sense that if $\qmx$ is a point at maximum depth then
$
d^-(q^-) \ge d_\eps^-(\qmx) \; \text{ and }\;
d^+(q^+) \ge d_{\eps/2}^-(\qmx) .
$
The running time is
$
\tilde{O}\left( \frac{\sqrt{n}}{\eps^{3/2}} + n + \frac{1}{\eps^2} \right) .
$
% \[
% O\left(
% \frac{\sqrt{n}}{\eps^{3/2}} \log^{1/2} (\eps n) + \left( n + \frac{1}{\eps^2} \right)
% \log\frac{1}{\eps}\right) .
% \]
\end{theorem}
%%%%%%%%%%%%%%%%%%%%%%%%%

%\medskip
%\noindent{\bf Remark.}
The naive approach to finding the maximum depth, that works only in the primal,
with the same $m=O(1/\eps^2)$ queries, takes
$O\left(\frac{n}{\eps} + \frac{1}{\eps^2}\log\frac{1}{\eps}\right)$ time.
Our solution is faster when $n =\tilde{\Omega} \left(\frac{1}{\eps}\right)$,
and the improvement becomes more significant as $n$ grows.
%Similar improvements arise in all the other subsequent cases, with the same threshold for $n$,
%but we will not state them explicitly in the other cases.

%%%%%%%%%%%%%%%%%%%%%%%%%%%%%%%%%%%%%%%%%%%%%%%%%%%%%
\section{Approximate depth for triangles} \label{sec:apxtri}
\label{sec:triangles}

In this section we  obtain
an efficient data structure for answering approximate depth queries
for triangles.
We avoid a multilevel structure in the dual by
decomposing each triangle into trapezoids.
This decomposition allows us to
reduce the problem into a problem on halfplanes before we even switch to the dual space.

%We will then use the structure to approximate the maximum depth.
%The underlying setup is rather similar to the one presented in Section~\ref{sec:apxhalf}
%for a collection of halfplanes, although there are some notable differences,
%discussed below. We actually reduce the case of triangles to
%the case of halfplanes, at the cost of a small increase in the performance bounds,
%by a logarithmic factor.

\begin{figure}[htb]
\begin{center}
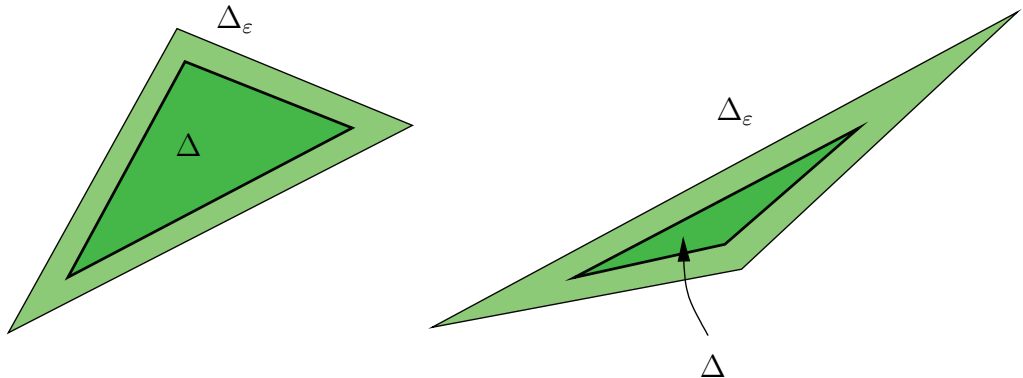
\caption{\sf The offset triangle $\Delta_\eps$ of a triangle $\Delta$.
In the right figure, the distance of a point $q\in\Delta_\eps$ from $\Delta$
can be much larger than $\eps$.}
\label{offset}
\end{center}
\end{figure}

Our input is a set $S$ of $n$ triangles, all contained in $Q=[0,1]^2$,
and an error parameter $\eps>0$. Given a query point $q$, the
\emph{inner $\eps$-depth} $d_\eps^-(q)$ of $q$ is the number of triangles
$\Delta$ in $S$ such that $\Delta$ contains $q$ and $q$ lies at distance
$\ge \eps$ from the boundary of $\Delta$, and the
\emph{outer $\eps$-depth} $d_\eps^+(q)$ of $q$ is the number of triangles
$\Delta\in S$ such that $q$ is contained in the offset triangle $\Delta_\eps$, whose edges
lie on the lines obtained by shifting each of the supporting lines of the
edges of $\Delta$ by $\eps$ away from $\Delta$; see Figure~\ref{offset}.
The reason for this somewhat different definition of $d_\eps^+(q)$ is
that the locus of points that are either contained in a
given triangle $\Delta$ or are at distance at most $\eps$ from its boundary,
which is the Minkowski sum of $\Delta$ with a disk of radius $\eps$, has `rounded
corners' bounded by circular arcs around the vertices of the triangle, and handling
such arcs does not work well in our duality-based approach 
(see Figure~\ref{fig:epsdepth}). Our modified definition avoids these circular arcs,
but it allows to include in $d_\eps^+(q)$ triangles $\Delta$ such that 
the distance of $q$ from $\bd\Delta$ is much larger than $\eps$
(see Figure~\ref{offset}(right)). Nevertheless, our technique
will avoid counting triangles with such an excessive deviation.

% As a matter of fact, we will
% estimate a somewhat smaller quantity---see below for details.

Our goal is to compute numbers $d^-(q)$ and $d^+(q)$ that satisfy \\
$d_\eps^-(q) \le d^-(q) \le d(q) \le d^+(q) \le d_\eps^+(q)$.

\subparagraph*{Reducing to the case of halfplanes.}
Let $\Delta$ be an arbitrary triangle. We represent $\Delta$ as the `signed union'
of three trapezoidal regions $R_1$, $R_2$, $R_3$, so that either
$\Delta = (R_1\cup R_2)\setminus R_3$, or $\Delta = R_3 \setminus (R_1\cup R_2)$,
and $R_1$ and $R_2$ are disjoint. To obtain these regions, we choose some
direction $u$ (details about the choice will be given shortly), and project
the three edges of $\Delta$ in direction $u$ onto a line $\ell_u^\bot$ orthogonal
to $u$ and lying outside $Q$. We say that an edge $e$ of $\Delta$ is \emph{positive}
(resp., \emph{negative}) in the direction $u$ if $e$ lies above (resp., below)
the interior of $\Delta$ in direction $u$, locally near $e$. To make $R_1$ and
$R_2$ disjoint, we make one of them half-open, removing from it the common vertical
edge that it shares with the other trapezoid. $\Delta$ has either two
positive edges and one negative edge, or two negative edges and one positive edge.
We associate with $e$ the trapezoid $R(e)$ whose bases are in direction $u$, one
of its side edges is $e$, and the other lies on $\ell_u^\bot$. $R(e)$
is \emph{positive} (resp., \emph{negative}) if $e$ is positive (resp., negative).

\begin{figure}[htb]
\begin{center}
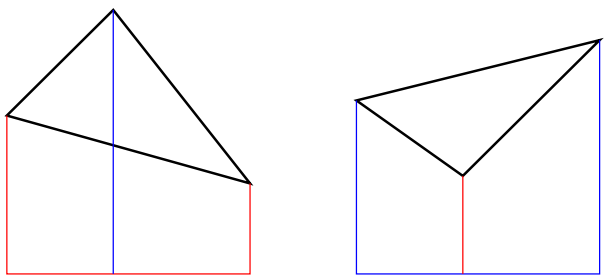
\caption{\sf Representing a triangle as the signed union of three trapezoids:
(a) The case of two positive edges and one negative edge.
(b) The case of two negative edges and one positive edge.}
\label{tri:right}
\end{center}
\end{figure}

Let $e_1$, $e_2$, $e_3$ be the three edges of $\Delta$, and denote $R(e_i)$
shortly as $R_i$, for $i=1,2,3$. It is clear from the construction that
$\Delta = (R_1\cup R_2)\setminus R_3$ when $e_1$ and $e_2$ are positive and
$e_3$ is negative, and $\Delta = R_3 \setminus (R_1\cup R_2)$ when $e_1$ and
$e_2$ are negative and $e_3$ is positive (one of these situations always holds
with a suitable permutation of the indices), and that $R_1$ and $R_2$ are
disjoint. See Figure~\ref{tri:right} for an illustration.
Moreover, the sum of the signs of the trapezoids that contain a point $q$
is $1$ if $q\in\Delta$ and $0$ otherwise.
% \micha{Actually, if we double $D$, we can always assume that we only have two
% positive edges and one negative edge. Not that it matters much.}

To control the distance of $q$ to the boundary of any triangle counted in
$d_\eps^+(q)$, we want to choose the direction $u$ so that the angles that $e_1$, $e_2$
and $e_3$ form with $u$ is at least some (large) positive angle $\beta$. 
(This will guarantee that the distance in direction $u$ of a point in
$\Delta_\eps\setminus \Delta$ from its nearest edge is at most some (small)
constant multiple of $\eps$.) The range of directions $u$
that violate this property for any single edge is at most $2\beta$,
so we are left with a range of good directions for $\Delta$ of size at least
$\pi-6\beta$. Hence, if $\beta$ is sufficiently smaller than $\pi/6$, we can
find a fixed set $D$ of $O(1)$ directions so that at least one of them will be
a good direction for $\Delta$, in the sense defined above.
Note that this choice of good directions is in fact a refinement of the argument
used in Section~\ref{sec:apxhalf} to control the slope of the lines bounding the
input halfplanes.

We assign each $\Delta\in S$ to one of its good directions in $D$, and construct,
for each $u\in D$, a separate data structure over the set $S_u$ of triangles
assigned to $u$. In what follows we fix one $u\in D$, assume without loss of
generality that $u$ is the positive $y$-direction, and
continue to denote by $S$ the set of triangles assigned to $u$.
We let $P$ and $N$ denote, respectively, the resulting sets of all
positive trapezoids and of all negative trapezoids.

We now construct a two-level data structure on the trapezoids in $P$ 
(the treatment of $N$ is fully symmetric). The first
level is a segment tree over the $x$-projections of the trapezoids of $P$.
For each node $v$ of the segment tree, let $P_v$ denote the set of trapezoids
of $P$ whose projections are stored at $v$. In what follows we can think
(for query points whose $x$-coordinate lies in the interval $I_v$ associated
with $v$) of each trapezoid $R\in P_v$ as a halfplane, bounded by the line
supporting the triangle edge that is the ceiling of $R$.

The storage and preprocessing cost of the segment tree are $O(n\log n)$, for $|S| = n$. 

At each node $v$ of the segment tree, the second level of the structure
at $v$ consists of an instance of the data structure of Section~\ref{sec:apxhalf},
constructed for the halfplanes associated with the trapezoids of $P_v$.\footnote{%
  Note that since we already did the slope partitioning globally for the triangles, 
  we do not need slope partitioning at the structure of the halfplanes.}

To answer a query with a point $q$, we search with $q$ in each of the $O(1)$
data structures, over all directions in $D$. For each direction, we search
separately in the `positive structure' and in the `negative structure'.
For the positive structure, we search with $q$ in the segment tree,
and for each of the $O(\log n)$ nodes $v$ that we reach, we access the
second-level structure of $v$ (constructed over the trapezoids of $P_v$),
and obtain the ($v$-dependent) counts $d^-(q)$, $d^+(q)$, which satisfy Equation (\ref{dplmi})
with respect to the halfplanes of the trapezoids in $P_v$. We sum up these quantities over 
all nodes $v$ on the search path of $q$. We do the same for the halfplanes of the trapezoids of $N_v$ for the same nodes $v$.

To avoid confusion we denote the relevant quantities of Equation (\ref{dplmi}) 
with respect to the union of the halfplanes of $P_v$ over all nodes $v$ in the search path of $q$ in the segment tree  as
$\pi_\eps^-(q)$, $\pi^-(q)$, $\pi(q)$, $\pi^+(q)$, and $\pi_\eps^+(q)$,
respectively. We denote the similar quantities for the union of the $N_v$'s as
$\nu_\eps^-(q)$, $\nu^-(q)$, $\nu(q)$, $\nu^+(q)$, and $\nu_\eps^+(q)$.

In summary, we have computed $\pi^-(q)$,
$\pi^+(q)$, and $\nu^-(q)$ and $\nu^+(q)$ such that
\begin{align} \label{ineq:pn}
\pi_\eps^-(q) & \le \pi^-(q) \le \pi(q) \le \pi^+(q) \le \pi_\eps^+(q) \\
\nu_\eps^-(q) & \le \nu^-(q) \le \nu(q) \le \nu^+(q) \le \nu_\eps^+(q) . \nonumber
\end{align}
We now set and output
\begin{equation} \label{dpinu}
d^-(q) := \pi^-(q) - \nu^+(q), \qquad\text{and}\qquad
d^+(q) := \pi^+(q) - \nu^-(q) .
\end{equation}

Recall that only
$\pi^-(q)$, $\pi^+(q)$, $\nu^-(q)$ and $\nu^+(q)$ depend on the
specific implementation of the structure, where the remaining values
are algorithm independent, depending only on $q$, $\eps$ and $P$ and $N$
(and on the set $D$ of directions and one the assignment of triangles
to directions).

%%%%%%%%%%%%%%%%%%%%%%
\begin{lemma} \label{lem:pn1}
We have, for any point $q\in Q$,
\[
d(q) = \pi(q) - \nu(q), \qquad d_\eps^-(q) = \pi_\eps^-(q) - \nu_\eps^+(q), \qquad\text{and} \qquad
d_\eps^+(q) = \pi_\eps^+(q) - \nu_\eps^-(q) .
\]
\end{lemma}
%%%%%%%%%%%%%%%%%%%%%%
%\noindent{\bf Proof.}
%See Appendix~\ref{app:apxtri}.
%$\Box$

\begin{figure}[htb]
\begin{center}
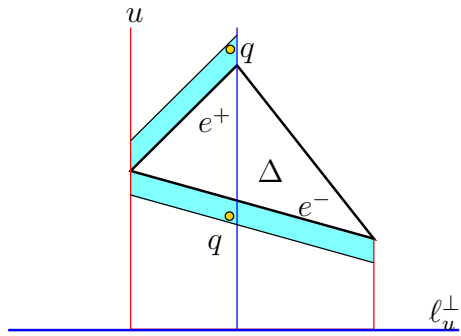
\caption{\sf The case where $q$ lies outside $\Delta$ but within distance
at most $\eps$ from a line supporting an edge ($e^+$ or $e^-$) of $\Delta$
(two such points $q$ are drawn).
$\Delta$ is counted in $\pi_\eps^+(q)$ but not in $\nu_\eps^-(q)$.}
\label{tri:pn}
\end{center}
\end{figure}

Using Lemma~\ref{lem:pn1} and the inequalities in (\ref{ineq:pn}), one easily obtains
the desired inequalities
$
d_\eps^-(q) \le d^-(q) \le d(q) \le d^+(q) \le d_\eps^+(q) .
$
% with the modified definition of $d_\eps^+(q)$.

The approximate maximum depth problem is handled as in Section~\ref{sec:apxhalf},
except that we use the $d^-$ and $d^+$ values as defined in (\ref{dpinu}).
Note that if a triangle $\Delta$ is counted in $d^+(q)$
(and $q$ lies outside $\Delta$) then the distance of $q$ from $\bd\Delta$ is at most
$\eps/\sin\beta$; this is our promised control of the distance deviation.
We thus obtain the following main results of this section.

%%%%%%%%%%%%%%%%%%%%%%%%%%
\begin{theorem}[Restatement of Theorem \ref{main:query} for triangles] \label{th:depth:tri}
Let $S$ be a set of $n$ triangles and let $\eps>0$ be an error parameter. We
can construct a data structure such that, for a query point $q$ in the unit square,
we can compute two numbers $d^-(q)$, $d^+(q)$ that satisfy
$
d_\eps^-(q) \le d^-(q) \le d(q) \le d^+(q) \le d_\eps^+(q) ,
$
with the modified definition of $d_\eps^+(q)$.
Denoting by $m$ the number of queries that we expect the structure to perform,
we can construct the structure so that its preprocessing cost and storage, and the time
it takes to answer $m$ queries, are both
$
\tilde{O}\left( \frac{\sqrt{mn}}{\sqrt{\eps}} + m + n \right) .
$
% \[
% O\left( \left(
% \frac{\sqrt{mn}}{\sqrt{\eps}}\log^{1/2} \frac{n}{\eps m}
% + \left( m + n \right) \log\frac{1}{\eps}\right) \log n\right) .
% \]
\end{theorem}
%%%%%%%%%%%%%%%%%%%%%%%%%%

%%%%%%%%%%%%%%%%%%%%%%%%%%
\begin{theorem} \label{th:maxdepth:tri}
Let $S$ be a set of $n$ triangles in the unit square, and let $\eps>0$ be an error parameter.
We can compute points $q^-$ and $q^+$ so that $d^-(q^-)$ and $d^+(q^+)$ closely approximate
the maximum depth in $\A(S)$, in the sense that if $\qmx$ is a point at maximum depth then
$d^-(q^-) \ge d_\eps^-(\qmx)$ and $d^+(q^+) \ge d_{\eps/2}^-(\qmx)$.
The running time is
$
\tilde{O}\left( \frac{\sqrt{n}}{\eps^{3/2}} + n + \frac{1}{\eps^2} \right) .
$
% \[
% O\left( \left(
% \frac{\sqrt{n}}{\eps^{3/2}} \log^{1/2} (\eps n) + \left( n + \frac{1}{\eps^2} \right)
% \log\frac{1}{\eps}\right) \log n \right).
% \]
\end{theorem}
%%%%%%%%%%%%%%%%%%%%%%%%%%

%%%%%%%%%%%%%%%%%%%%%%%%%%%%%%%%%%%%
\section{Approximate depth for halfspaces in higher dimensions} \label{sec:apxhalfs}
\label{sec:halfspaces}

The technique in Section~\ref{sec:apxhalf} can easily be extended to any higher dimension
$d\ge 3$. Due to lack of space we only state our results here and refer the reader to
Appendix~\ref{app:apxhalfs} for full details.
Here we have a set $S$ of $n$ halfspaces in $\reals^d$, whose bounding
hyperplanes cross the unit cube $Q = [0,1]^d$, and an error parameter $\eps>0$,
and we want to preprocess $S$ into a data structure that allows us to answer
approximate depth queries efficiently for points in $Q$, as well as to find
a point in $Q$ of approximate maximum depth, where both tasks are qualified
as in Section~\ref{sec:apxhalf}. The high-level approach is a fairly
natural generalization of the techniques in Section~\ref{sec:apxhalf},
albeit quite a few of the steps of the extension are technically nontrivial,
and require some careful calculations and calibrations of the relevant parameters.
Our results are:

%%%%%%%%%%%%%%%%%%%%%%%%%%
\begin{theorem} \label{th:depthd}
Let $S$ be a set of $n$ halfspaces in $\reals^d$ and let $\eps>0$ be an error parameter.
We can construct a data structure such that, for a query point $q$ in the unit cube
$[0,1]^d$, we can compute two numbers $d^-(q)$, $d^+(q)$ that satisfy
$
d_\eps^-(q) \le d^-(q) \le d(q) \le d^+(q) \le d_\eps^+(q) .
$
Denoting by $m$ the number of queries that we expect the structure to perform,
we can construct the structure so that its preprocessing cost and storage, and the time
it takes to answer $m$ queries, are all
$
\tilde{O}\left( \frac{\sqrt{mn}}{\eps^{(d-1)/2}} + n + m \right) .
$
% \[
% O\left( \frac{\sqrt{mn}}{\eps^{(d-1)/2}} \log^{1/2}\frac{n}{\eps^{d-1} m}
% + n \log\frac{1}{\eps} + m \log\frac{1}{\eps}\right) .
% \]
\end{theorem}

%%%%%%%%%%%%%%%%%%%%%%%%%%
\begin{theorem} \label{th:maxdepthd}
Let $S$ be a set of $n$ halfspaces in $\reals^d$ and let $\eps>0$ be an error parameter. We
can compute points $q^-$ and $q^+$ so that $d^-(q^-)$ and $d^+(q^+)$ closely approximate the maximum depth
in $\A(S)$ within $[0,1]^d$, in the sense that if $\qmx$ is a point at maximum depth then
$d^-(q^-) \ge d_\eps^-(\qmx)$ and $d^+(q^+) \ge d_{\eps/2}^-(\qmx)$. The running time is
$
\tilde{O}\left( \frac{\sqrt{n}}{\eps^{d-1/2}} + n + \frac{1}{\eps^d} \right) .
$
\end{theorem}

Our technique is faster than the naive bound
$O\left( \frac{n}{\eps^{d-1}} + \frac{1}{\eps^d}\log\frac{1}{\eps}\right)$
when $n = \tilde{\Omega} \left( \frac{1}{\eps}\right)$.

%%%%%%%%%%%%%%%%%%%%%%%%%%%%%%%%%%%%
\section{Approximate depth for simplices in higher dimensions} \label{sec:apxsimp}
\label{sec:simplices}

The results of Section~\ref{sec:apxtri} can be extended to higher dimensions.
To simplify the presentation, we describe, in Appendix~\ref{app:apxsimp}, the case
of three dimensions in detail, and then comment on the extension to any higher dimension.
here we only consider the general $\reals^d$ case. 

% \subparagraph*{Simplices in three dimensions.}
%%%%%%%%%%%%%%%%%%%%%%%%%%%%%%
\begin{theorem} \label{simp3d}
Let $S$ be a set of $n$ simplices within the unit cube $Q = [0,1]^d$ in $\reals^d$,
and let $\eps>0$ be an error parameter. We can construct a data structure on $S$
which computes, for any query point $q\in Q$, an underestimate $d^-(q)$ and an
overestimate $d^+(q)$ on the depth of $q$ in $S$, which satisfy
$
d^-_\eps(q) \le d^-(q) \le d(q) \le d^+(q) \le d^+_\eps(q) ,
$
under the modified definition of $d_\eps^+(q)$ (as presented in the introduction).
If $m$ is the number of queries that we want or expect to perform, the preprocessing
cost of the structure, and the time to answer $m$ queries, are both
${\displaystyle
\tilde{O}\left( \frac{m^{2/(d+2)}n^{d/(d+2)}}{\eps^{d(d-1)/(d+2)}} + m + n \right) 
}$.
\end{theorem}
%%%%%%%%%%%%%%%%%%%%%%%%%%%%%%
Note that the dependence on $\eps$ is better than in the naive solution, as
$d(d-1)/(d+2) < d-1$. % The improvement diminishes as $d$ gets larger.

%%%%%%%%%%%%%%%%%%%%%%%%%%%%%%
\begin{theorem} \label{simp3dmx}
Let $S$ be a set of $n$ simplices in the unit cube $Q = [0,1]^d$ in $\reals^d$,
and let $\eps>0$ be an error parameter.
We can compute points $q^-$ and $q^+$ so that $d^-(q^-)$ and $d^+(q^+)$ closely approximate
the maximum depth in $\A(S)$, in the sense that if $\qmx$ is a point at maximum depth then
$d^-(q^-) \ge d_\eps^-(\qmx)$ and $d^+(q^+) \ge d_{\eps/2}^-(\qmx)$. The running time is
\[
\tilde{O}\left( \frac{n^{d/(d+2)}}{\eps^{d(d+1)/(d+2)}} + n + \frac{1}{\eps^d} \right) .
\]
\end{theorem}
%%%%%%%%%%%%%%%%%%%%%%%%%%%%%%

Here too, this beats the naive solution when $n =\tilde{\Omega} \left( \frac{1}{\eps} \right)$.

%
% %%%%%%%%%%%%%%%%%%%%%%%%%%%%%%%%%%%%%%%%%%%%%%%%%%%%%
% \section{Discussion} \label{sec:disc}
%
% \micha{Care to add a discussion / open problems / whatever?}

\bibliography{grid}

\begin{thebibliography}{10}

\bibitem{AC09}
Peyman Afshani and Timothy~M. Chan.
\newblock On approximate range counting and depth.
\newblock {\em Discrete Comput. Geom.}, 42(1):3--21, 2009.

\bibitem{Ag}
Pankaj Agarwal.
\newblock Simplex range searching and its variants: A review.
\newblock In M.~Loebl, J.~Ne\v{s}et\v{r}il, and R.~Thomas, editors, {\em A
  Journey through Discrete Mathematics: A Tribute to Jiri Matousek}, pages
  1--30. Springer Verlag, Berlin-Heidelberg, 2017.

\bibitem{AE}
Pankaj~K. Agarwal and Jeff Erickson.
\newblock Geometric range searching and its relatives.
\newblock In B.~Chazelle, J.E. Goodman, and R.~Pollack, editors, {\em Advances
  in Discrete and Computational Geometry}, pages 1--56. AMS Press, Providence
  RI, 1998.

\bibitem{AKKSZ}
Dror Aiger, Haim Kaplan, Efi Kokiopoulou, Micha Sharir, and Bernhard Zeisl.
\newblock General techniques for approximate incidences and their application
  to the camera posing problem.
\newblock In {\em Proc. 35th Internat. Sympos. Comput. Geom.}, pages 8:1--8:14,
  2019.
\newblock Also in arXiv:1903.07047.

\bibitem{AKS}
Dror Aiger, Haim Kaplan, and Micha Sharir.
\newblock Output sensitive algorithms for approximate incidences and their
  applications.
\newblock In {\em Proc. European Sympos. Algorithms}, pages 1--13, 2017.

\bibitem{AHP}
Boris Aronov and Sariel Har{-}Peled.
\newblock On approximating the depth and related problems.
\newblock {\em {SIAM} J. Comput.}, 38(3):899--921, 2008.

\bibitem{AM00}
Sunil Arya and David~M. Mount.
\newblock Approximate range searching.
\newblock {\em Comput. Geom.}, 17(3-4):135--152, 2000.

\bibitem{RANSAC}
Robert~C. Bolles and Martin~A. Fischler.
\newblock A {RANSAC}-based approach to model fitting and its application to
  finding cylinders in range data.
\newblock In {\em Proc. 7th Internat. Joint Conf. Artificial Intelligence},
  pages 637--643, 1981.

\bibitem{Breuel2003}
Thomas~M. Breuel.
\newblock Implementation techniques for geometric branch-and-bound matching
  methods.
\newblock {\em Computer Vision and Image Understanding}, 90(3):258--294, 2003.

\bibitem{chan1996}
Timothy~M. Chan.
\newblock Fixed-dimensional linear programming queries made easy.
\newblock In {\em Proc. 12th ACM Sympos. Comput. Geom.}, pages 284--290, 1996.

\bibitem{FM10}
Guilherme~D. Da~Fonseca and David~M. Mount.
\newblock Approximate range searching: The absolute model.
\newblock {\em Comput. Geom.}, 43(4):434--444, 2010.

\bibitem{FLO2015}
Johan Fredriksson, Viktor Larsson, and Carl Olsson.
\newblock Practical robust two-view translation estimation.
\newblock In {\em Proc. IEEE Conference on Computer Vision and Pattern
  Recognition}, pages 2684--2690, 2015.

\bibitem{FLO2016}
Johan Fredriksson, Viktor Larsson, Carl Olsson, and Fredrik Kahl.
\newblock Optimal relative pose with unknown correspondences.
\newblock In {\em Proc. IEEE Conference on Computer Vision and Pattern
  Recognition}, pages 1728--1736, 2016.

\bibitem{HarPeled:book}
Sariel Har-Peled.
\newblock {\em Geometric Approximation Algorithms}.
\newblock AMS Press, Providence RI, 2011.

\bibitem{HW04}
Sariel Har-Peled and Yusu Wang.
\newblock Shape fitting with outliers.
\newblock {\em SIAM J. Comput.}, 33(2):269--285, 2004.

\bibitem{Haussler1987}
David Haussler and Emo Welzl.
\newblock $\eps$-nets and simplex range queries.
\newblock {\em Discrete Comput. Geom.}, 2(2):127--151, 1987.

\bibitem{KaplanRS11}
Haim Kaplan, Edgar Ramos, and Micha Sharir.
\newblock Range minima queries with respect to a random permutation, and
  approximate range counting.
\newblock {\em Discrete Comput. Geom.}, 45(1):3--33, 2011.

\bibitem{OKO2009}
Carl Olsson, Fredrik Kahl, and Magnus Oskarsson.
\newblock Branch and bound methods for euclidean registration problems.
\newblock {\em IEEE Trans. Pattern Anal. Mach. Intell.}, 31(5):783--794, 2009.

\end{thebibliography}

\appendix

The appendices below contain the full version of the technical parts of the paper, with a few minor modifications.

%%%%%%%%%%%%%%%%%%%%%%%%%%%%%%%%%%%%
\section{Approximate depth for halfplanes} \label{app:apxhalf}

To illustrate our approach, we begin with the simple case where $S$ is a collection of
$n$ halfplanes. We construct a data structure that computes numbers $d^-(q)$, $d^+(q)$
that satisfy (\ref{dplmi}), for queries $q$ in the square $Q=[0,1]^2$, and for some
prespecified error parameter $\eps>0$.
We denote by $\ell_h$ the boundary line of a halfplane $h\in S$.

We remark that this simplest setup, of halfplanes in $\reals^2$, is treated in a manner
that is similar to the algorithms for computing $\eps$-incidences in previous work~\cite{AKS}.
We spell it out in detail because it is simpler to present, and helps to set the stage
or the more involved cases of triangles in $\reals^2$ and of halfsplaces and simplices in
higher dimensions, where the results presented here (i) are considerably different, (ii) need
a battery of additional technical steps, (iii) yield substantially improved solutions (when $n$
is reasonably large in terms of $\eps$ and when the number of queries is not too excessive),
and (iv) are in fact novel, as the depth problem, under the fuzzy model assumed here
(and in \cite{AM00}), does not seem to have been considered in the previous works.
Although the depth problem is, in a sense, a dual variant of the range counting problem,
it raises new technical challenges, which do not have matching counterparts in the range
searching context, and addressing these challenges is far from trivial, as we will
demonstrate in these appendices (and, in part, also in the main part of the paper).

A straightforward way to do this is to construct an (uncompressed) quadtree $T$ within
$Q$ in the standard manner. For $i\ge 0$, let $T^i$ denote the $i$-th level of $T$.
Thus $T^0$ consists of $Q$ as a single square, $T^1$ consists of four $1/2\times 1/2$
subsquares, and in general $T^i$ consists of $4^i$ subsquares of side length $1/2^i$.
For technical reasons, it is advantageous to have the squares at each level pairwise
disjoint, and we ensure this by making them half-open. Concretely, each square $\tau$
is of the form $a\le x<a+\delta$, $b\le y<b+\delta$, where $(a,b)$ is the vertex of $\tau$
with smallest coordinates and $\delta$ is the side length of $\tau$. This holds for most
squares, except that the rightmost squares are also closed on their right side and the
topmost squares are also closed on their top side.

We stop the construction when we reach the level $k = \lceil \log (\sqrt{2}/\eps) \rceil$,
so the diameter of the squares of $T^k$ is at most $\eps$.
We denote by $\tau_v$ the square represented by a node
$v\in T$, and by $p(v)$ the parent of $v$ in $T$. We construct a pruned version of
$T$ in which each node $v$, such that no line $\ell_h$ crosses $\tau_v$, but there is
at least one line $\ell_h$ that crosses $\tau_{p(v)}$, becomes a
`shallow' leaf and is not expanded further.

For each node $v$ of $T$ (other than the root), we maintain a counter $c(v)$ of the number of halfplanes
$h$ that fully contain $\tau_v$ but are such that $\ell_h$ crosses the parent square
$\tau_{p(v)}$ of $\tau_v$, and for each `deep' leaf $v$, at the bottom level $T^k$,
we maintain an additional counter $b(v)$ of the number of halfplanes $h$ whose
boundary lines $\ell_h$ cross $\tau_v$. (The shallow leaves have $b(v)=0$,
as some deep leaves might also have.)

We construct $T$ by incrementally inserting into it each $h\in S$
(creating nodes on the fly as needed), as follows.
Initially, $T$ consists of just $Q$ itself, with a $c$-value of $0$. When the
insertion of $h$ reaches level $i$, we have already updated the counters of all
the relevant nodes at levels $\le i$, and we have constructed a list $E_i$ of the
nodes at level $i$ that $\ell_h$ crosses. We check the containment / crossing relation
of the four children of each $v\in E_i$ with respect to $h$ and $\ell_h$, increment
$c(w)$ for each child $w$ of $v$ such that $\tau_w$ is contained in $h$, and insert
into $E_{i+1}$ each child $w$ of $v$ such that $\ell_h$ crosses $\tau_w$.
(The insertion of $h$ may cause nodes $v$ that so far have been shallow leaves
to be expanded in $T$ into their four child squares.)
We start this insertion process by initializing $E_0$ to contain the root.
We wrap up the process by incrementing $b(v)$ for each node $v$ in $E_k$.
At the end of the process, we mark all the unexpanded nodes at levels shallower
than $k$ as \emph{shallow} leaves, and set their $b$-counters to $0$.
Note that, by construction, the process never expands such a leaf.

To answer a query $q$, we set $d^-(q)$ to be the sum of the counters $c(v)$ of all nodes
$v$ on the path to the leaf $u$ containing $q$, and set $d^+(q):= d^-(q)+b(u)$.
Note that both values $d^-(q)$ and $d^+(q)$ depend only on the leaf $v$ containing $q$.
We denote these values also as $d^-(v)$ and $d^+(v)$, respectively, and note that they
can be computed during the construction of $T$ at no extra asymptotic cost.

The correctness of this algorithm is easy to establish. For $d^-(q)$, we clearly only
count halfplanes that fully contain $q$. Moreover, the boundary line of any halfplane
counted in $d_\eps^-(q)$ cannot cross the bottom-level square that contains $q$, so
it will get counted in the $c$-counter of (exactly) one of the squares that we visit.
Hence $d_\eps^-(q) \le d^-(q) \le d(q)$, for any query point $q$.

For $d^+(q)$, if $q\in h$ for some $h\in S$, we will either count $h$ in the $c$-counter
of one of the squares that we visit, up to the leaf square containing $q$, inclusive,
or count $h$ in the $b$-counter of the leaf. Moreover, the boundary line of each
halfplane that we count at the $b$-counter of the leaf must be at distance at most
$\eps$ from $q$. Hence $d(q) \le d^+(q) \le d_\eps^+(q)$, for any query point $q$.
These observations establish the correctness of the procedure.

As is easily verified, the time needed to construct (the pruned) $T$ and to compute
the counters of its nodes is $O\left(\frac{n}{\eps} \right)$.
% \micha{Mumble something about the model of computation? Using here the floor function?}
The time it takes to answer a query with a point $q$ is
$O\left(\log\frac{1}{\eps}\right)$: All we need to do is to find the leaf $v$ containing $q$
and retrieve the precomputed values $d^-(v)$ and $d^+(v)$.
We can use our data structure to compute a leaf $v\in T$ of maximum $d^-(v)$,
and a (possibly different) leaf $v$ of maximum $d^+(v)$, by simply iterating over all leaves,
in time proportional to the size of $T$ (which is at most $O(1/\eps^2)$).
The maximum value of $d^-(v)$ (resp., of $d^+(v)$)
is an underestimate (resp., overestimate) of the maximum depth in $\A(S)$.
While these numbers can vary significantly from the maximum depth itself, such a discrepancy
is caused solely by ``near-containments'' (false or shallow) of a point in many halfplanes.
The same holds for the depth of an arbitrary query point $q$, in the sense that the
possible discrepancy between $d^-(q)$ and $d(q)$, and between $d(q)$ and $d^+(q)$,
are caused only by shallow and false containments of $q$, respectively.
We note that when the input has some inaccuracy or uncertainty, up to a
displacement by $\eps$, the actual depth of a point $q$ can assume any value
between $d^-_\eps(q)$ and $d^+_\eps(q)$.

%%%%%%%%%%%%%%%%%%%%%%%%%%%%%%%%%%%%%%%%%%%%%%%%
\subsection{Faster construction using duality} \label{app:halfplanes-duality}

We now show how to use duality to improve the storage and preprocessing cost of this
data structure, at the expense of larger query time. We will then balance these costs
to obtain a more efficient procedure for answering many queries and, consequently,
also for approximate maximum depth. We use standard duality
that maps each point $p=(\xi,\eta)$ to the line $p^*:\; y=\xi x-\eta$, and each line
$\ell:\;y=cx+d$ to the point $\ell^*=(c,-d)$. This duality preserves the vertical
distance $\ddv$ between the point and the line; that is, $\ddv(p,\ell) = \ddv(\ell^*,p^*)$.

Since duality preserves the vertical distance and not the standard distance, we want
the vertical distance to be a good approximation of the actual distance. This is not true
in general, but we can ensure this by restricting the slope of the input boundary lines,
as we describe below.

We construct our primal quadtree $T$ within $Q$ exactly as before, making its cells
half-open as described above, but this time only
up to level $k = \log \frac{1}{\delta_1}$, for some parameter $\eps\le \delta_1 \le 1$.
Now each leaf $v$ in $T^k$ represents a square $\tau_v$ of side length $\delta_1$.
(We ignore the rather straightforward rounding issues in what follows, or simply
insist that $\delta_1$ (and $\eps$ too) be a negative power of $2$.) We compute
the counters $c(v)$ for all nodes $v\in T$, but we do not need the counters $b(v)$
for the $k$-level leaves of $T$.  Instead, for each deep leaf $v\in T^k$,
we pass to the dual plane and construct there a dual quadtree on the set of
points dual to the boundary lines that cross $\tau_v$.
(Only leaves at the bottom level require this dual construction.)

Let $\tau=\tau_v$ be a square associated with some bottom-level leaf $v$ of $T^k$.
Let $S_\tau \subseteq S$ be the subset of halfplanes $h$ whose boundary line
$\ell_h$ crosses $\tau$. We partition the halfplanes in $S_\tau$ into four subsets
according to the slope of their boundary lines.\footnote{%
  We can increase the number of such subsets, thereby improving the distortion
  between the real and vertical distances.}
Each family, after an appropriate
rotation, consists only of halfplanes whose boundary lines have slopes in $[0,1]$.
We focus on the subset where the original boundary lines have slope in $[0,1]$,
and abuse the notation slightly by denoting it as $S_\tau$ from now on. The treatment
of the other subsets is analogous. The input to the corresponding dual problem at $\tau$
is the set $S_\tau^*$ of points dual to the boundary lines of the halfplanes in $S_\tau$.
In general, each $\tau$ has four subproblems associated with it.

We assume without loss of generality that $\tau = [0,\delta_1]^2$.
It follows from our slope condition that the boundary lines of the halfplanes in
$S_\tau$ intersect the $y$-axis in the interval $[-\delta_1,\delta_1]$.
Therefore, by the definition of the duality transformation, each dual point
$h^* \in S_\tau^*$ lies in the rectangle $R_\tau^* = [0,1]\times [-\delta_1,\delta_1]$.
Any square other than $\tau$ is treated analogously, except that the duality has to
be adjusted. If $\tau = [a\delta_1,(a+1)\delta_1]\times [b\delta_1,(b+1)\delta_1]$
then we modify the duality so that we first move $\tau$ to $[0,\delta_1]^2$, and
then apply the standard duality.\footnote{%
  Concretely, as is easily verified, we map each point $q\in\tau$ to the line
  $q^*:\; y = (q_x-a)x-(q_y-b)$, and map each line $\ell:\; y = cx+d$ to the
  point $\ell^* = (c,-d-ac+b)$.}

We store the points of $S_\tau^*$ in a dual pruned quadtree $T_\tau$, whose root
corresponds to $R_\tau^*$, and for each $i$, its $i$-th level $T^i_\tau$
corresponds to a partition of $R_\tau^*$ into $2^i \times 2^i$ congruent rectangles,
each of side lengths $(1/2^i) \times (2\delta_1/2^i)$. We stop the construction
when we reach level $k^* = \log \frac{4}{\delta_2}$, for another parameter
$\delta_2$, also assumed to be a suitable negative power of $2$, at which
each rectangle associated with a leaf $u$ is of width $\delta_2/4$ and of height
$\delta_1 \delta_2/2$. We constrain the choice of $\delta_1$ and $\delta_2$
by requiring that $\delta_1 \delta_2 = \eps$ (so, as already mentioned, we
assume here that $\eps$ is also a negative power of $2$).

Consider a query point $q\in \tau$ and let $q^*$ be its dual line. Let $h$ be a
halfplane in $S_\tau$ and let $h^*$ be its dual point (that is, the point dual to
its boundary line). Now $q$ lies in $h$ if and only if $h^*$ lies in an appropriate
side of $q^*$: this is the upper (resp., lower) side if $h$ is an upper (resp., lower)
halfplane. We therefore encode the direction (upper/lower) of $h$ with $h^*$, by
defining $h^*$ to be \emph{positive} if $h$ is an upper halfplane and \emph{negative}
if $h$ is a lower halfplane. Each node $u$ of $T_\tau$ stores two counters $c^+(u)$
and $c^-(u)$ of the positive and negative points, respectively, of $S_\tau^*$ that
are contained in the rectangle represented by $u$.

We answer a query with a point $q$ as follows (consult Figure~\ref{sketch}). 
We first search the primal quadtree
$T$ for the leaf $v$ such that $q\in \tau_v$. If $v$ is a shallow leaf,
we stop the process and output the sum of the counters $c(u)$ over all nodes $u$
on the search path to $v$, inclusive; note that in this case we obtain the real depth of $q$.
Otherwise, we search in the dual quadtree $T_{\tau_v}$ with the line $q^*$, and
sum the counts $c^+(u)$ of all nodes $u$ whose rectangle lies above $q^*$ but
the rectangle of the parent of $u$ is crossed by $q^*$, and the counts $c^-(u)$
of all nodes $u$ whose rectangle lies below $q^*$ but the rectangle of the parent
of $u$ is crossed by $q^*$. We denote by $C^-(q)$ and $C^+(q)$ these two respective
sums. Let $C(v)$ be the sum of the counters $c(u)$ in the primary tree of all nodes
$u$ along the path from the root to $v$. We set $d^-(q) := C(v) + C^-(q) + C^+(q)$,
and set $d^+(q)$ to be $d^-(q)$ plus the sum of all the counters $c^+(u)+c^-(u)$
of the leaves $u$ of $T_{\tau_v}$ that $q^*$ crosses.

\subparagraph*{Correctness.}
The correctness of this procedure is argued as follows.
%%%%%%%%%%%%%%%%%%%%%%%%%%%%%%%%
\begin{lemma} \label{hs-correct}
(a) For any query point $q$ we have $d^-(q) \le d(q) \le d^+(q)$. \\
(b) Let $h$ be a halfplane in $S$.
If $q$ lies in $h$ at distance larger than $\eps$ from $\ell_h$ then $h$ is counted in $d^-(q)$. \\
(c) If $h$ is counted in $d^+(q)$ then the distance between $q$ and $h$ is at most $\eps$.
\end{lemma}
%%%%%%%%%%%%%%%%%%%%%%%%%%%%%%%%
\noindent{\bf Proof.}
Part (a) follows easily from the construction, using a similar reasoning to that in the
primal-only approach presented above. For part (b),
assume without loss of generality that $q\in \tau = [0,\delta_1]^2$.
Assume that $q$ lies in $h$ at distance larger than $\eps$ from $\ell_h$.
If $q$ lies in a primal square $\tau_u$ that $\ell_h$ misses but crosses its parent square,
then we count $h$ in $c(u)$, and thus in $d^-(q)$ (the specific assumption made in (b)
is not used here). Otherwise, $\ell_h$ must cross the primal
leaf square $\tau_v$ that contains $q$, and then $h^*$ appears in the dual subproblem
associated with $\tau=\tau_v$. Again, if we reach some dual node $u$ whose rectangle
contains $h^*$, is missed by $q^*$, and lies on the correct side of $q^*$,
we count $h$ in either $c^+(u)$ or $c^-(u)$ (overall, we count
$h$ at most once in this manner). Otherwise $q^*$ would have to cross the rectangle of
the bottom-level leaf $u$ of $T_\tau$ that contains $h^*$. This however is impossible.
Indeed, we have $\eps \le \dd(q,\ell_h) \le \ddv(q,\ell_h) = \ddv(h^*,q^*)$.
Since $q\in \tau$, the slope of $q^*$ is between $0$ and $\delta_1$.
Furthermore, the width and height of the dual rectangle at $u$ are $\delta_2/4$ and
$\delta_1\delta_2/2$, respectively. Thus $q^*$ is at vertical distance at least
\[
\eps - \frac{\delta_1\delta_2}{4} - \frac{\delta_1\delta_2}{2} = \frac{\eps}{4}
\]
from any point in the dual rectangle, and in particular $q^*$ does not
intersect that rectangle, as claimed. It follows that we count $h^*$ in (exactly)
one of the counters $c^-(u)$ or $c^+(u)$, over the proper ancestors of the
secondary leaf containing $h^*$. In either of the above cases,
$h^*$ is counted in $d^-(q)$.

\begin{figure}[htb]
\begin{center}
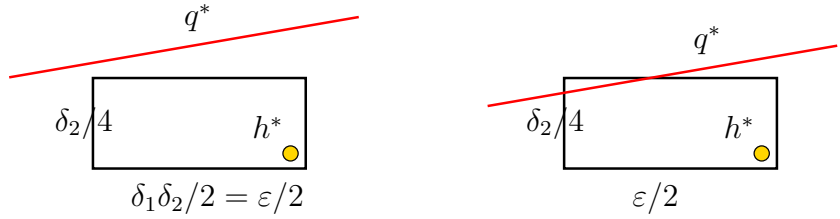
\caption{\sf Illustration of the proof of Lemma~\ref{hs-correct}.
Left: The argument for part (b).  Right: The argument for part (c).}
\label{dualeaf}
\end{center}
\end{figure}

Similarly, for part (c) of the lemma,
either we count $h$ in a counter $c(u)$ of some primal node $u$ whose
square $\tau_u$ contains $q$ and is fully contained in $h$ (and then $q\in h$
for sure), or else $\ell_h$ crosses the $k$-level primal leaf square
$\tau=\tau_v$ that contains $q$, and then we count $h$ in one of the dual subproblems
at $\tau$. Indeed, this happens either when we count $h$ in some node $u$ of $T_\tau$
that contains $h^*$ and is missed by $q^*$ (and then again $q\in h$ for sure),
or else we count $h$ in the $c^+$ or $c^-$ counters of the secondary leaf $u$ at the bottom-level
$k^*$ of $T_\tau$ whose dual rectangle contains $h^*$. In this case $q^*$ crosses this
rectangle. Assuming, as above, that $\tau_v = [0,\delta_1]^2$, the slope of
$q^*$ is in $[0,\delta_1]$. This, and the fact that $q^*$ crosses the rectangle
containing $h^*$, imply that the vertical distance from $h^*$ to $q^*$ is at most
\[
\frac{\delta_1\delta_2}{2} + \frac{\delta_1\delta_2}{4} = \frac{3\eps}{4} < \eps .
\]
Hence, the vertical distance from $q$ to $h$ is at most $\eps$, and therefore
so is the real distance from $q$ to $h$, as claimed.
$\Box$

\subparagraph*{Preprocessing and storage.}
Suppressing the expansion of the primal quadtree at nodes that are not crossed by
any boundary line makes the storage that it requires $O\left(\frac{n}{\delta_1}\right)$,
and it can be constructed in time $O\left(\frac{n}{\delta_1}\right)$.
Fix a primal bottom-level leaf square $\tau=\tau_v$, and put $n_\tau := |S_\tau|$.
It takes $O\left(n_\tau \log \frac{1}{\delta_2}\right)$ time and space to construct $T_\tau$.
(Similar to the primal setup, we prune $T_\tau$ so as not to explicitly represent nodes whose
rectangles do not contain any dual point.) Since we have $\sum_\tau n_\tau = O(n/\delta_1)$,
over all $k$-level leaf squares $\tau$ of the primal tree $T$, we get that the total construction time
of all dual structures is $O\left(\frac{n}{\delta_1} \log \frac{1}{\delta_2}\right)$,
and this also bounds their overall storage. Together, the total construction time
and storage is therefore $O\left(\frac{n}{\delta_1} \log \frac{1}{\delta_2}\right)$.

When we answer a query $q$, it takes $O\left(\log \frac{1}{\delta_1}\right)$ time to
find the leaf $v$ in $T$ whose square $\tau_v$ contains $q$, and then,
assuming $v$ to be a bottom-level leaf,
$O\left(\frac{1}{\delta_2}\right)$ time to trace $q^*$ in
$T_{\tau_v}$ and add up the appropriate counters. The total cost of a query is thus
\[
O\left( \frac{1}{\delta_2} + \log \frac{1}{\delta_1} \right) ,
\]
and the total time for $m$ queries is
${\displaystyle O\left(m \left(\frac{1}{\delta_2} + \log \frac{1}{\delta_1}\right)\right)}$.
It is easy to see that the term $\log\frac{1}{\delta_1}$ dominates only when $\delta_2$
is very close to $1$. Specifically this happens when
${\displaystyle \frac{1}{\log\frac{1}{\eps}} \le \delta_2 \le 1}$.

\subparagraph*{Analysis.}
Let $m$ denote the number of queries that we want (or expect) to handle.
The values of $\delta_1$ and $\delta_2$ that nearly balance the construction time with the
total time for $m$ queries, under the constraint that $\delta_1\delta_2=\eps$, are
(ignoring the issue of possible dominance of the term $\log\frac{1}{\delta_1}$
in the query cost)
\[
\delta_1 = \sqrt{\frac{n\eps}{m}} ,\quad\quad
\delta_2 = \sqrt{\frac{m\eps}{n}} ,
\]
% \[
% \delta_1 = \sqrt{\frac{n\eps}{m}} \sqrt{\log \frac{n}{\eps m}} ,\quad\quad
% \delta_2 = \sqrt{\frac{m\eps}{n}}\frac{1}{\sqrt{\log \frac{n}{\eps m}}} ,
% \]
and the cost is then
\begin{equation} \label{eq:bd1}
\tilde{O}\left( \frac{\sqrt{mn}}{\sqrt{\eps}} \right) .
\end{equation}
For this to make sense, we must have $\eps \le \delta_1,\;\delta_2 \le 1$,
which holds when
\[
n\eps \le m \le \frac{n}{\eps} ,
\]
% Equivalently, putting $x=\eps m/n$, we require
% \[
% \eps^2 \log\frac{1}{x} \le x \le \log\frac{1}{x} ,
% \]
which means that
\[
% c_1\eps^2\log\frac{1}{\eps} \le x \le c_2 ,\qquad\text{or}\qquad
c_1\eps n \le m \le c_2 n ,
\]
for suitable absolute constants $c_1$, $c_2$.
Note that when $m$ is close to the upper bound of this range,
$\log\frac{1}{\delta_1}$, which is then $\log\frac{1}{\eps}$,
dominates $\frac{1}{\delta_2}$, and the overall cost of the queries becomes
$O(m\log\frac{1}{\eps}) = \tilde{O}(m)$, a term that will appear later in the overall
bound anyway.

In this range, this bound is better than the
naive bound of $O\left(\frac{n}{\eps}+m\log\frac{1}{\eps}\right)$ yielded
by our naive fully primal solution, and is also better than the bound
$O\left(\frac{m}{\eps}+n\log\frac{1}{\eps}\right)$ that we would obtain if we
applied the naive scheme only in the dual.
When $m< c_1\eps n$, % \log\frac{n}{\eps m}$, 
we only work in the dual, for a cost of
\begin{equation} \label{eq:bd2}
O\left(\frac{m}{\eps}+n\log\frac{1}{\eps}\right) = \tilde{O}\left( n \right) ,
\end{equation}
and when $m>\frac{c_2n}{\eps}$, we only work in the primal plane, for a cost of
\begin{equation} \label{eq:bd3}
O\left(\frac{n}{\eps}+m\log\frac{1}{\eps}\right) = \tilde{O}\left( m \right) .
\end{equation}
Hence the total cost of $m$ queries, including the preprocessing cost, results
by adding the bounds in (\ref{eq:bd1}), (\ref{eq:bd2}), and (\ref{eq:bd3}), and is
\begin{equation} \label{eq:bd}
\tilde{O}\left( \frac{\sqrt{mn}}{\sqrt{\eps}} + n + m \right) .
\end{equation}

The following theorem summarizes our result.
%%%%%%%%%%%%%%%%%%%%%%%%%%%%%%55
\begin{theorem}[Restatement of Theorem \ref{main:query} for halfplanes] \label{th:depth}
    Let $S$ be a set of $n$ halfplanes in $\reals^2$ and let $\eps>0$ be an error parameter.
    We can construct a data structure such that, for a query point $q$, we can compute
    two numbers $d^-(q)$, $d^+(q)$ that satisfy
    \[
    d_\eps^-(q) \le d^-(q) \le d(q) \le d^+(q) \le d_\eps^+(q) .
    \]
    Denoting by $m$ the number of queries that we expect the structure to perform,
    we can construct the structure so that its preprocessing cost and storage, and the time
    it takes to answer $m$ queries, are both
    \[
    \tilde{O}\left( \frac{\sqrt{mn}}{\sqrt{\eps}} + m + n \right) .
    \]
    % \[
    % O\left(
    % \frac{\sqrt{mn}}{\sqrt{\eps}}\log^{1/2} \frac{n}{\eps m} + \left( m + n \right)
    % \log\frac{1}{\eps}\right) .
    % \]
    % The storage required by the structure is
    % \[
    % O\left(
    % \frac{\sqrt{mn}}{\sqrt{\eps}}\log^{1/2} \frac{n}{\eps m} + \left( m + n \right)
    % \log\frac{1}{\eps}\right).
    % \]
\end{theorem}
%%%%%%%%%%%%%%%%%%%%%%%%%%%%%%55

%%%%%%%%%%%%%%%%%%%%%%%%%%%%%%%%%%%%%%%%%%%%%%
\subparagraph*{Approximating the maximum depth.}
We can use our data structure to approximate the maximum depth as follows.
For each primal $\frac{\eps}{2\sqrt{2}} \times \frac{\eps}{2\sqrt{2}}$
grid square $\sigma$, pick its center $q_\sigma$, compute
$d^-(q_\sigma)$ and $d^+(q_\sigma)$, using our structure,
and report $d^-=\max_\sigma d^-(q_\sigma)$ and $d^+=\max_\sigma d^+(q_\sigma)$
(and, if desired, also the squares attaining these maxima).
%More precisely, for each query point $q_\sigma$ we compute and add up four sub-counts
%for each of $d^-(q_\sigma)$, $d^+(q_\sigma)$, one for each subset of input halfspaces
%whose boundary lines have slopes in a corresponding restricted range of slopes.
%\haim{Unnecessary repetition in my opinion}
In this application the number of queries is $m = O\left( 1/\eps^2 \right)$.

Lemma \ref{half-correct} and Theorem \ref{th:maxdepth}
that follow
specify  the properties of $d^-$ and $d^+$.

%asserts, in a precise specific manner, the correctness of this
%procedure. It essentially shows that it suffices to compute the depth of the
%center points of the grid squares, since the depth of any other point $q$, or
%rather the underestimate $d^-(q)$ and the overestimate $d^+(q)$,
%do not deviate much from the corresponding estimates of an adjacent center point.

%%%%%%%%%%%%%%%%%%%%%%%%%
\begin{lemma} \label{half-correct}
(a) Let $q$ be an arbitrary point in $Q$, and let $\sigma$ be the
$\frac{\eps}{2\sqrt{2}} \times \frac{\eps}{2\sqrt{2}}$ grid square
that contains $q$. Then we have
\[
d_\eps^-(q)  \le d^-(q_\sigma) \;\text{ and }\;
d_{\eps/2}^-(q)  \le d^+(q_\sigma) \ .
\]

\noindent
(b) In particular, let $\qmx$ be a point of maximum (exact) depth in $\A(S)$,
and let $\sigma$ be the
$\frac{\eps}{2\sqrt{2}} \times \frac{\eps}{2\sqrt{2}}$ grid square
that contains $\qmx$. Then we have
\[
d_\eps^-(\qmx)  \le d^-(q_\sigma) \; \text{ and } \;
d_{\eps/2}^-(\qmx)  \le d^+(q_\sigma) \ .
\]
\end{lemma}
%%%%%%%%%%%%%%%%%%%%%%%%%
\noindent{\bf Proof.}
We only prove (a), since (b) is just a special case of it.
We establish each inequality separately.

\noindent{\bf (i)}
$d_\eps^-(q) \le d^-(q_\sigma)$:
Let $h$ be a halfplane that contains $q$, so that $q$ lies at distance
greater than $\eps$ from $\ell_h$.
Since the distance between $q$ and $q_\sigma$
is $\le \eps/4$ then  $q_\sigma$ is also in $h$.

We claim that $h$ must be counted in $d^-(q_\sigma)$ before we reach a leaf either in the primal or the dual processing.

We prove this claim by contradiction as follows
If we do not count $h$ in $d^-(q_\sigma)$
during the primal and dual processing then
it must be the case that $q_\sigma^*$ crosses the bottom-level dual rectangle that
contains $h^*$.  As in the proof of Lemma~\ref{hs-correct}(c), this implies that
the vertical distance between $q_\sigma$ and $\ell_h$ is at most $\frac34\eps$.
But then it follows that
 the distance from $q$ to
$\ell_h$ is at most $\eps$, a contradiction.

\noindent{\bf (ii)}
$d_{\eps/2}^-(q) \le d^+(q_\sigma)$:
Let $h$ be a halfplane that contains $q$ and $q$ lies at distance
at least $\eps/2$ from $\ell_h$. In this case it is is also clear that $h$ contains
$q_\sigma$, so $h$ will be counted in $d^+(q_\sigma)$ by the preceding arguments.
No assumption on the grid size is needed in this case.
$\Box$

This lemma implies the following.
%%%%%%%%%%%%%%%%%%%%%%%%%%%%%%%%%%%
\begin{theorem}[Restatement of Theorem \ref{th:maxdepth-main}] \label{th:maxdepth}
Let $S$ be a set of $n$ halfplanes in $\reals^2$ and let $\eps>0$ be an error parameter.
We can compute points $q^-$ and $q^+$ such that $d^-(q^-)$ and $d^+(q+)$ closely approximate
the maximum depth in $\A(S)$, in the sense that if $\qmx$ is a point at maximum depth then
\[
d^-(q^-) \ge d_\eps^-(\qmx) \; \text{ and }\;
d^+(q^+) \ge d_{\eps/2}^-(\qmx) .
\]
The running time is
\[
\tilde{O}\left( \frac{\sqrt{n}}{\eps^{3/2}} + n + \frac{1}{\eps^2} \right) .
\]
% \[
% O\left(
% \frac{\sqrt{n}}{\eps^{3/2}} \log^{1/2} (\eps n) + \left( n + \frac{1}{\eps^2} \right)
% \log\frac{1}{\eps}\right) .
% \]
\end{theorem}
%%%%%%%%%%%%%%%%%%%%%%%%%%

\noindent{\bf Remarks.}
{\bf (a)}
The bound in Theorem \ref{th:depth} is smaller than the bound obtained from
(a suitable adaptation of) Arya and Mount's bound~\cite{AM00}, which is
$O\left(m\log\frac{1}{\eps} + \frac{n}{\eps}\right)$,
when $m = \tilde{O}\left( \frac{n}{\eps}\right)$ (otherwise, both bounds are
$\tilde{O}\left(m\right)$). Similarly, the bound in
Theorem \ref{th:maxdepth} is better than the bound in \cite{AM00} when
$n = \tilde{\Omega}\left( \frac{1}{\eps}\right)$.

\medskip
\noindent{\bf (b)}
As already discussed, a priori,
in both parts of the theorem, the output counts $d^-$, $d^+$ (or $d^-(q^-)$, $d^+(q^+)$) could vary significantly
from the actual depth $d(q)$ (or maximum depth). Nevertheless, such a discrepancy is caused
only because the query point (or the point of maximum depth) lies too close to the
boundaries (either inside or outside) of many halfplanes in $S$.

%%%%%%%%%%%%%%%%%%%%%%%%%%%%%%%%%%%%%%%%%%%%%%%%%%%%%
\section{Approximate depth for triangles} \label{app:apxtri}

In this section we adapt the approach used in Appendix~\ref{app:apxhalf} to obtain
an efficient data structure for answering approximate depth queries
for triangles. We will then use the structure to approximate the maximum depth.
Our technique is to reduce the case of triangles to
the case of halfplanes by
decomposing the triangles into trapezoids.
This allows us to avoid the need for a multilevel structure in the dual space.

% at the cost of a small increase in the performance bounds,
%by a logarithmic factor.

Our input is a set $S$ of $n$ triangles, all contained in, or more generally
overlap $Q=[0,1]^2$,
and an error parameter $\eps>0$. Given a query point $q$, the
\emph{inner $\eps$-depth} $d_\eps^-(q)$ of $q$ is the number of triangles
$\Delta$ in $S$ such that $\Delta$ contains $q$ and $q$ lies at distance
at least $\eps$ from the boundary of $\Delta$, and the
\emph{outer $\eps$-depth} $d_\eps^+(q)$ of $q$ is the number of triangles
$\Delta\in S$ such that the `offset' triangle $\Delta_\eps$, whose edges
lie on the lines obtained by shifting each of the supporting lines of the
edges of $\Delta$ by $\eps$ away from $\Delta$; see Figure~\ref{offset}.

% \begin{figure}[htb]
% \begin{center}
% \input{offset.pstex_t}
% \caption{\sf The offset triangle $\Delta_\eps$ of a triangle $\Delta$.
% In the right figure, the distance of a point $q\in\Delta_\eps$ from $\Delta$
% can be much larger than $\eps$.}
% \label{offset}
% \end{center}
% \end{figure}

As a matter of fact, we will estimate a somewhat smaller quantity,
to control the effect of sharp corners in (the offset of) $\Delta$, which may
be too far from $\bd \Delta$---see below for details.
Our goal is to compute numbers $d^-(q)$ and $d^+(q)$ that satisfy
\[
d_\eps^-(q) \le d^-(q) \le d(q) \le d^+(q) \le d_\eps^+(q) .
\]
The reason for this somewhat different definition of $d_\eps^+(q)$ comes
from the fact that the locus of points that are either contained in a
given triangle $\Delta$ or are at distance at most $\eps$ from its boundary,
which is the Minkowski sum of $\Delta$ with a disk of radius $\eps$, has `rounded
corners' bounded by circular arcs around the vertices of the triangle, and handling
such arcs does not work well in a duality-based approach, like ours
(see Figure~\ref{fig:epsdepth}). Our modified definition avoids these circular arcs,
but it may include triangles $\Delta$ that are included in $d_\eps^+(q)$ even though
the distance of $q$ from $\bd\Delta$ is much larger than $\eps$. Our technique
will avoid counting triangles with such an excessive deviation.

\subparagraph*{Reducing to the case of halfplanes.}
Let $\Delta$ be an arbitrary triangle. We represent $\Delta$ as the `signed union'
of three trapezoidal regions $R_1$, $R_2$, $R_3$, so that either
$\Delta = (R_1\cup R_2)\setminus R_3$, or $\Delta = R_3 \setminus (R_1\cup R_2)$,
and $R_1$ and $R_2$ are disjoint. To obtain these regions, we choose some
direction $u$ (details about the choice will be given shortly), and project
the three edges of $\Delta$ in direction $u$ onto a line $\ell_u^\bot$ orthogonal
to $u$ and lying outside $Q$. We say that an edge $e$ of $\Delta$ is \emph{positive}
(resp., \emph{negative}) in the direction $u$ if $e$ lies above (resp., below)
the interior of $\Delta$ in direction $u$, locally near $e$. To make $R_1$ and
$R_2$ disjoint, we make one of them half-open, removing from it the common vertical
edge that it shares with the other trapezoid. $\Delta$ has either two
positive edges and one negative edge, or two negative edges and one positive edge.
We associate with $e$ the trapezoid $R(e)$ whose bases are in direction $u$, one
of its side edges is $e$, and the other lies on $\ell_u^\bot$. We say that $R(e)$
is \emph{positive} (resp., \emph{negative}) if $e$ is positive (resp., negative).

Let $e_1$, $e_2$, $e_3$ be the three edges of $\Delta$, and denote $R(e_i)$
shortly as $R_i$, for $i=1,2,3$. It is clear from the construction that
$\Delta = (R_1\cup R_2)\setminus R_3$ when $e_1$ and $e_2$ are positive and
$e_3$ is negative, and $\Delta = R_3 \setminus (R_1\cup R_2)$ when $e_1$ and
$e_2$ are negative and $e_3$ is positive (one of these situations always holds
with a suitable permutation of the indices), and that $R_1$ and $R_2$ are
disjoint. See Figure~\ref{tri:right} for an illustration.
Moreover, the sum of the signs of the trapezoids that contain a point $q$
is $1$ if $q\in\Delta$ and $0$ otherwise.
% \micha{Actually, if we double $D$, we can always assume that we only have two
% positive edges and one negative edge. Not that it matters much.}

% \begin{figure}[htb]
% \begin{center}
% \input{posneg12.pstex_t}
% \caption{\sf Representing a triangle as the signed union of three trapezoids:
% (a) The case of two positive edges and one negative edge.
% (b) The case of two negative edges and one positive edge.}
% \label{tri:right}
% \end{center}
% \end{figure}

To control the distance of $q$ to the boundary of any triangle counted in
$d_\eps^+(q)$, we want to choose the direction $u$ so that none of the angles that $e_1$, $e_2$
and $e_3$ form with $u$ is too small; concretely, we want each of these angles
to be at least some (large) positive angle $\beta$. The range of directions $u$
that violate this property for any single edge is at most $2\beta$,
so we are left with a range of good directions for $\Delta$ of size at least
$\pi-6\beta$. Hence, if $\beta$ is sufficiently smaller than $\pi/6$, we can
find a fixed set $D$ of $O(1)$ directions so that at least one of them will be
a good direction for $\Delta$, in the sense defined above.
Note that this choice of good directions is in fact a refinement of the argument
used in Appendix~\ref{app:apxhalf} to control the slope of the lines bounding the
input halfplanes.

We assign each $\Delta\in S$ to one of its good directions in $D$, and construct,
for each $u\in D$, a separate data structure over the set $S_u$ of triangles
assigned to $u$. In what follows we fix one $u\in D$, assume without loss of
generality that $u$ is the positive $y$-direction, and
continue to denote by $S$ the set of triangles assigned to $u$.
We let $P$ and $N$ denote, respectively, the resulting sets of all
positive trapezoids and of all negative trapezoids.

We now construct a two-level data structure on the trapezoids in $P$. The first
level is a segment tree over the $x$-projections of the trapezoids of $P$.
For each node $v$ of the segment tree, let $P_v$ denote the set of trapezoids
of $P$ whose projections are stored at $v$. In what follows we can think
(for query points whose $x$-coordinate lies in the interval $I_v$ associated
with $v$) of each trapezoid $R\in P_v$ as a halfplane, bounded by the line
supporting the triangle edge that is the ceiling of $R$.

The storage and preprocessing cost of the segment tree are $O(n\log n)$,
for an input set of $n$ triangles.

At each node $v$ of the segment tree, the second level of the structure
at $v$ consists of an instance of the data structure of Appendix~\ref{app:apxhalf},
constructed for the halfplanes associated with the trapezoids of $P_v$.\footnote{Note that since
    we already did the slope partitioning globally for the triangles, we do not need slope partitioning at the structure of the halfplanes.}

To answer a query with a point $q$, we search with $q$ in each of the $O(1)$
data structures, over all directions in $D$. For each direction, we search
separately in the `positive structure' and in the `negative structure'.
For the positive structure, we search with $q$ in the segment tree,
and for each of the $O(\log n)$ nodes $v$ that we reach, we access the
second-level structure of $v$ (constructed over the trapezoids of $P_v$),
and obtain the ($v$-dependent) counts $d^-(q)$, $d^+(q)$, which satisfy Equation (\ref{dplmi})
with respect to the halfplanes of the trapezoids in $P_v$. We sum up these quantities over all nodes $v$ on the search path of $q$. We do the same for the halfplanes of the trapezoids of $N_v$ for the same nodes $v$.

To avoid confusion we
denote the relevant quantities of Equation (\ref{dplmi}) with respect to the union of the halfplanes of $P_v$ over all nodes $v$ in the search path of $q$ in the segment tree  as
$\pi_\eps^-(q)$, $\pi^-(q)$, $\pi(q)$, $\pi^+(q)$, and $\pi_\eps^+(q)$,
respectively. We denote the similar quantities for the union of the $N_v$'s
as
$\nu_\eps^-(q)$, $\nu^-(q)$, $\nu(q)$, $\nu^+(q)$, and $\nu_\eps^+(q)$.

In summary, we have computed $\pi^-(q)$,
$\pi^+(q)$, and $\nu^-(q)$ and $\nu^+(q)$ such that
\begin{align} \label{app:ineq:pn}
\pi_\eps^-(q) & \le \pi^-(q) \le \pi(q) \le \pi^+(q) \le \pi_\eps^+(q) \\
\nu_\eps^-(q) & \le \nu^-(q) \le \nu(q) \le \nu^+(q) \le \nu_\eps^+(q) . \nonumber
\end{align}
We now set and output
\begin{equation} \label{app:dpinu}
d^-(q) := \pi^-(q) - \nu^+(q), \qquad\text{and}\qquad
d^+(q) := \pi^+(q) - \nu^-(q) .
\end{equation}

Recall that $\pi^-(q)$, $\pi^+(q)$, $\nu^-(q)$ and $\nu^+(q)$ depend on the
specific implementation of the structure, where the remaining values
are algorithm independent, depending only on $q$, $\eps$ and $P$ and $N$
(and on the set $D$ of directions and the assignment of triangles
to directions).

%%%%%%%%%%%%%%%%%%%%%%
\begin{lemma} \label{app:lem:pn1}
We have, for any point $q\in Q$,
\[
d(q) = \pi(q) - \nu(q), \qquad d_\eps^-(q) = \pi_\eps^-(q) - \nu_\eps^+(q), \qquad\text{and} \qquad
d_\eps^+(q) = \pi_\eps^+(q) - \nu_\eps^-(q) .
\]
\end{lemma}
%%%%%%%%%%%%%%%%%%%%%%
\noindent{\bf Proof.}
The first identity is immediate from the construction.

For the second identity, let $\Delta$ be a triangle that contains $q$ so that
$q$ lies at distance at least $\eps$ from $\bd\Delta$. As is easily checked,
this is equivalent to the property that $q$ lies at distance at least $\eps$
from each of the three lines supporting the edges of $\Delta$, on the side
of that line that contains $\Delta$. Let $e^+$ and $e^-$ be the edges of
$\Delta$ that lie above and below $q$ (in the appropriate direction $u$), respectively. Then $e^+\in P$ and $e^-\in N$.
By the definition of $\pi_\eps^-(q)$ and
$\nu_\eps^+(q)$, $\Delta$ contributes $+1$
to $\pi_\eps^-(q)$ but is not counted in
$\nu_\eps^+(q)$.
The converse direction is proved analogously.

For the third identity assume that  $q$ lies  in the `offset' triangle
of $\Delta$.
% Assume further that, for the direction
%$u$ associated with $\Delta$, $q$ lies within the projection of $\Delta$
%onto $\ell_u^\bot$.
Let $e^+$ and $e^-$ be the edges of  $\Delta$
whose 'offset' edges
 lie above and below $q$ (in the appropriate direction $u$), respectively, so
$e^+\in P$ and $e^-\in N$.
 Now $q$ lies either slightly above $e^+$ or slightly
below $e^-$. In either case,
by the definition of $\pi_\eps^+(q)$ and
$\nu_\eps^-(q)$, $\Delta$ is counted in
$\pi_\eps^+(q)$ but not in
$\nu_\eps^-(q)$. The converse direction is proved analogously.
$\Box$

\medskip

Using Lemma~\ref{app:lem:pn1} and the inequalities in (\ref{app:ineq:pn}), one easily obtains
the desired inequalities
\[
d_\eps^-(q) \le d^-(q) \le d(q) \le d^+(q) \le d_\eps^+(q) ,
\]
with the modified definition of $d_\eps^+(q)$.

The approximate maximum depth problem is handled as in Appendix~\ref{app:apxhalf},
except that we use the $d^-$ and $d^+$ values as defined in (\ref{app:dpinu}).
Note that if a triangle $\Delta$ is counted in $d^+(q)$
(and $q$ lies outside $\Delta$) then the distance of $q$ from $\bd\Delta$ is at most
$\eps/\sin\beta$.

We thus obtain the summary results of this section, as stated as Theorems
\ref{th:depth:tri} and \ref{th:maxdepth:tri} in the main part of the paper.

%%%%%%%%%%%%%%%%%%%%%%%%%%%%%%%%%%%%
\section{Approximate depth for halfspaces in higher dimensions} \label{app:apxhalfs}

The technique in Section~\ref{sec:apxhalf} (and Appendix~\ref{app:apxhalf})
can easily be extended to any higher dimension
$d\ge 3$. Here we have a set $S$ of $n$ halfspaces in $\reals^d$, whose bounding
hyperplanes cross the unit cube $Q = [0,1]^d$, and an error parameter $\eps>0$,
and we want to preprocess $S$ into a data structure that allows us to answer
approximate depth queries efficiently for points in $Q$, as well as to find
points in $Q$ of approximate maximum depth, where both tasks are qualified
as in Section~\ref{sec:apxhalf}.

The high-level approach is a fairly straightforward generalization of the
techniques in Section~\ref{sec:apxhalf}. Nevertheless, at the risk of
some redundancy, we spell out its details
to some extent, because quite a few of the steps of the extension are
technically nontrivial, and require some careful calculations and
calibrations of the relevant parameters, and because of the
various applications, that are more meaningful in higher dimensions,
as mentioned in the introduction.

We use the same standard duality that maps each point $p=(\xi_1,\ldots,\xi_d)$
to the hyperplane $p^*:\; x_d = \sum_{k=1}^{d-1} \xi_k x_k - \xi_d$, and each hyperplane
$h:\; x_d = \sum_{k=1}^{d-1} \eta_k x_k - \eta_d$ to the point $h^* = (\eta_1,\ldots,\eta_d)$.
As in the planar case, this duality preserves the vertical distance $\ddv$
(in the $x_d$-direction) between the point and the hyperplane;
that is, $\ddv(p,h) = \ddv(h^*,p^*)$.

Again, since duality preserves the vertical distance and not the standard distance, we want
the vertical distance to be a good approximation of the actual distance. This is not true
in general, but we ensure this by restricting the normal directions of the input
boundary hyperplanes (normalized to unit vectors) to lie in a suitable small neighborhood
within the unit sphere,
combined with a suitable rotation of the coordinate frame.

More precisely, we partition the halfspaces in $S$ into $O(1)$ subsets, so that the
inward unit normals to hyperplanes in a subset (namely normals that point
into the input halfspace bounded by the hyperplane) all lie in some cap
of $\sph^{d-1}$ of opening angle at most $\varphi$, for some sufficiently
small constant parameter $\varphi$ that we will fix shortly.
For each such cap, we rotate the coordinate frame so that the center
of the cap lies in the positive $x_d$-direction (at the so-called `north pole'
of $\sph^{d-1}$). It is then easy to check that, for any point $p$ and
any hyperplane with normal direction in that cap, we have
\begin{equation} \label{cap:ineq}
\dd(p,h) \le \ddv(p,h) \le
\frac{\dd(p,h)}{\cos\varphi} \approx \left(1+\frac12\varphi^2\right) \dd(p,h) .
\end{equation}
We continue the presentation for a single such subset, and simplify
the notation by continuing to refer to it as $S$, and assume
that the center of the corresponding cap is on the positive $x_d$-axis
(so no rotation is needed).

We construct a primal octree $T$ within $Q$, similar to the quadtree construction
in the plane, making its cells half-open as in Section~\ref{sec:apxhalf}, up to level
$k = \log \frac{1}{\delta_1}$, for some parameter $\eps\le \delta_1 \le 1$.
Nodes that are not crossed by any bounding hyperplane become (shallow) leaves of the tree.
Now each leaf $v$ in the bottommost level $T^k$ represents a cube $\tau_v$
of side length $\delta_1$. We compute counters $c(v)$, for all nodes $v\in T$,
defined as the number of halfspaces that contain $\tau_v$ but do not contain
the cube at the parent of $v$. For each `deep' leaf $v\in T^k$,
we pass to the dual $\reals^d$ and construct there a dual octree on the set of
points dual to the boundary hyperplanes that cross $\tau_v$.
% where each point is equipped with a sign that specifies which side of
% it (above or below) corresponds to the primal halfspace that it encodes.
(Only leaves at the bottom level require this dual construction.)

Let $\tau=\tau_v$ be a cube associated with some bottom-level leaf $v$ of $T^k$.
Let $S_\tau \subseteq S$ be the subset of halfspaces $h\in S$ whose boundary
hyperplane $\bd h$ crosses $\tau$ (and has inward normal in the cap).
Before continuing, we note that the partition of $S$ into the ``cap subsets''
is not really needed in the primal part of the structure, but only in the
dual part, which we are about to discuss. Nevertheless, to simplify the
presentation, we apply this partition for the entire set $S$ at the
beginning of the preprocessing, and end up with
$O(1/\varphi^{d-1}) = O(1)$ subproblems, one for each cap.
The preprocessing and querying procedures have to be repeated these many times,
but in what follows we only consider one such subset, and, as mentioned,
continue to denote it as $S$.

The input to the corresponding dual problem at $\tau$ is the set $S_\tau^*$
of points dual to the boundary hyperplanes of the halfspaces in $S_\tau$.

We assume without loss of generality that $\tau = [0,\delta_1]^d$.
%(See the discussion in Section~\ref{sec:apxhalf} for the way to handle other
%cubes (squares there), an appropriate extension of which applies here too.)
By construction, the inward unit normal vectors to the boundary hyperplanes
of the halfspaces in $S_\tau$ all lie in the $\varphi$-cap $C_\varphi$
of $\sph^{d-1}$ centered at the $x_d$-unit vector $e_d = (0,\ldots,0,1)$.
In the notation introduced earlier, the equation of any halfspace $h\in S_\tau$
is of the form $x_d \ge \sum_{i=1}^{d-1} \eta_i x_i - \eta_d$, so that
the corresponding inward normal unit vector is
\[
\nn_h = \frac{(-\eta_1,\ldots,-\eta_{d-1},1)}{\sqrt{1+\|\eeta\|^2}} ,
\]
where $\eeta = (\eta_1,\ldots,\eta_{d-1})$ and the norm is the Euclidean norm.
Since $\nn_h\in C_\varphi$, we have
\[
1 \ge \nn_h\cdot e_d = \frac{(-\eta_1,\ldots,-\eta_{d-1},1)\cdot e_d}{\sqrt{1+\|\eeta\|^2}} \ge\cos\varphi ,
\]
or
\[
\frac{1}{\sqrt{1+\|\eeta\|^2}} \ge\cos\varphi ,\qquad\text{or}\qquad \|\eeta\| \le\tan\varphi .
\]
Since the hyperplane bounding $h$, given by $x_d = \sum_{i=1}^{d-1} \eta_i x_i - \eta_d$,
crosses $\tau$, there are vertices $(x_1,\ldots,x_d)\in\{0,\delta_1\}^d$ of $\tau$
that lie above the hyperplane and vertices that lie below it. This is easily
seen to imply that
\[
- \left(1 + \|\eeta\|_1\right) \delta_1 \le \eta_d \le \|\eeta\|_1 \delta_1 ,
\]
where $\|\eeta\|_1$ is the $L_1$-norm of $\eeta$. By the Cauchy-Schwarz inequality,
we have
\[
\|\eeta\|_1 \le (d-1)^{1/2}\|\eeta\| \le (d-1)^{1/2}\tan\varphi .
\]
Choosing $\varphi$ so that
$(d-1)^{1/2}\tan\varphi = 1$, we have $-2\delta_1 \le \eta_d \le \delta_1$.
% \micha{Is this the best choice for $\varphi$?}

Therefore, by the definition of the duality transformation, each dual point
$h^* \in S_\tau^*$ lies in the Cartesian product
$R_\tau^* = B_{d-1}(0,\tan\varphi) \times [-2\delta_1,\delta_1]$, where
$B_{d-1}(0,\tan\varphi)$ is the $(d-1)$-dimensional ball of radius
$\tan\varphi$ centered at the origin. To simplify matters, we replace
$B_{d-1}(0,\tan\varphi)$ by the containing cube
\[
Q_\varphi = [-\tan\varphi,\tan\varphi]^{d-1} =
\left[-\frac{1}{(d-1)^{1/2}}, \frac{1}{(d-1)^{1/2}}\right]^{d-1} .
\]
We accordingly replace $R_\tau^*$ by $Q_\varphi \times [-2\delta_1,\delta_1]$.
As mentioned above, and elaborated in Section~\ref{sec:apxhalf} (for the planar case),
any cube other than $\tau$ is treated analogously, with a suitable coordinate shift.

We store the points of $S_\tau^*$ in a dual pruned octree $T_\tau$, whose root
corresponds to $R_\tau^*$, and for each $i\ge 0$, its $i$-th level $T^i_\tau$
corresponds to a partition of $R_\tau^*$ into $2^{id}$ congruent boxes,
each of side lengths $\displaystyle{ \frac{1}{\sqrt{d-1}2^{i-1}} \times \cdots \times
\frac{1}{\sqrt{d-1}2^{i-1}} \times \frac{3\delta_1}{2^i}}$. We stop the construction
when we reach level
\[
k^* = \log \frac{2}{\beta\delta_2} ,\qquad\text{for}\qquad
\beta = \frac{1}{\frac43 \sqrt{d-1}+2} ,
\]
for another parameter $\delta_2$, also assumed to be a suitable negative
power of $2$, at which each box associated with a leaf $u$ is of side lengths
\[
\frac{\beta\delta_2}{\sqrt{d-1}} \times \cdots \times
\frac{\beta\delta_2}{\sqrt{d-1}} \times \frac{3\beta\delta_1\delta_2}{2} .
\]

We constrain the choice of $\delta_1$ and $\delta_2$
by requiring that $\delta_1 \delta_2 = \eps$ (so, as already mentioned, we
assume here that $\eps$ is also a negative power of $2$).

Consider a query point $q\in \tau$ and let $q^*$ be its dual hyperplane. Let $h$ be a
halfspace in $S_\tau$ and let $h^*$ be its dual point (that is, the point dual to
its boundary hyperplane). Now $q$ lies in $h$ if and only if $h^*$ lies in an appropriate
side of $q^*$ (by our conventions, this is the upper side).
Each node $u$ of $T_\tau$ stores a counter $c^*(u)$ of the points of $S_\tau^*$ that
are contained in the box represented by $u$.

We answer a query with a point $q$ as follows (consult Figure~\ref{sketch}. 
We repeat what follows for each
of the $O(1)$ caps that cover $\sph^{d-2}$. We first search the primal octree
$T$ for the leaf $v$ such that $q\in \tau_v$. If $v$ is a shallow leaf,
we stop the process and output the sum of the counters $c(u)$ over all nodes $u$
on the search path to $v$, inclusive; note that in this case we obtain the real depth of $q$.
Otherwise (i.e., $v\in T^k$), we search in the dual octree $T_{\tau_v}$ with the
hyperplane $q^*$, and sum the counts $c^*(u)$ of all nodes $u$ whose box lies
above $q^*$ but the box of the parent of $u$ is crossed by $q^*$ (these nodes
are suitable children of the nodes encountered during the search).
We denote by $C^*(q)$ the resulting sum. Let $C(v)$ be the sum of the
counters $c(u)$ of all nodes $u$ in the primal tree
along the path from the root to $v$. We set $d^-(q) := C(v) + C^*(q)$,
and set $d^+(q)$ to be $d^-(q)$ plus the sum of all the counters $c^*(u)$
of the leaves $u$ of $T_{\tau_v}$ that $q^*$ crosses.
The actual values $d^-(q)$ and $d^+(q)$ that we return are the
sums of these quantities over all the caps.

\subparagraph*{Correctness.}
The correctness of this procedure is argued as in the planar case, except that
various sizes and other parameters have changed by suitable constant factors.
%%%%%%%%%%%%%%%%%%%%%%%%%%%%%%%%
\begin{lemma} \label{hs-correctd}
(a) For any query point $q$ we have $d^-(q) \le d(q) \le d^+(q)$. \\
(b) Let $h$ be a halfspace in $S$.
If $q$ lies in $h$ at distance larger than $\eps$ from $\bd h$ then $h$ is counted in $d^-(q)$. \\
(c) If $h$ is counted in $d^+(q)$ then the distance between $q$ and $h$ is at most $\eps$.
\end{lemma}
%%%%%%%%%%%%%%%%%%%%%%%%%%%%%%%%

\medskip
\noindent{\bf Proof.}
Part (a) is argued exactly as in the planar case. For part (b),
assume without loss of generality that $q\in \tau = [0,\delta_1]^d$
(recall the previous discussions concerning this issue).
Assume that $q$ lies in $h$ at distance larger than $\eps$ from $\bd h$.
If $q$ lies in a primal cube $\tau_u$ that $\bd h$ misses but crosses its parent cube,
then we count $h$ in $c(u)$, and thus in $d^-(q)$ (here we only need to assume
that $q\in h$). Otherwise, $\bd h$ must cross the primal
leaf cube $\tau_v$ that contains $q$, and then $h^*$ appears in the dual subproblem at
$\tau=\tau_v$. Again, if we reach some dual node $u$ whose box contains $h^*$,
is missed by $q^*$ (but its parent box is met by $q^*$),
and lies on the correct (that is, upper) side of $q^*$,
we count $h$ in $c^*(u)$ (overall, we count $h$ at most once in this manner).
Otherwise $q^*$ crosses the box of the bottom-level leaf $u$ of $T_\tau$
that contains $h^*$. This however is impossible. Indeed, if $q=(q_1,\ldots,q_d)$,
the equation of $q^*$ is $x_d = \sum_{k=1}^{d-1} q_kx_k - q_d$, and, by assumption,
this hyperplane meets the box $\tau^*_u$ of dimensions
\[
\frac{\beta\delta_2}{\sqrt{d-1}} \times \cdots \times
\frac{\beta\delta_2}{\sqrt{d-1}} \times \frac{3\beta\delta_1\delta_2}{2} .
\]
Hence, the maximum vertical distance, in the $x_d$-direction, of $q^*$
from $h^*$ (which lies in this box) is at most
\[
\frac{3\beta\delta_1\delta_2}{2} +
\sum_{k=1}^{d-1} \frac{\beta\delta_2}{\sqrt{d-1}} |q_k| .
\]
Since $q\in [0,\delta_1]^d$, this is at most
\[
\frac{3\beta\delta_1\delta_2}{2} +
(d-1)\delta_1 \cdot \frac{\beta\delta_2}{\sqrt{d-1}} =
\frac{1}{\frac43\sqrt{d-1}+2} \left(
\frac{3\delta_1\delta_2}{2} + \sqrt{d-1}\delta_1 \delta_2 \right) = \frac34 \delta_1\delta_2 = \frac34 \eps ,
\]
so the actual distance between $q^*$ and $h^*$ is also at most $\frac{3}{4}\eps$,
contradicting our assumption. It follows that we count $h^*$ in (exactly) one of the
counters $c(u)$ or $c^*(u)$. In either of the above cases, $h^*$ is counted in $d^-(q)$.

Similarly, for part (c) of the lemma,
either we count $h$ in a counter $c(u)$ of some primal node $u$ whose
cube $\tau_u$ contains $q$ and is fully contained in $h$ (and then $q\in h$
for sure), or else $\bd h$ crosses the $k$-level primal leaf cube
$\tau=\tau_v$ that contains $q$, and then we count $h$ in one of the dual subproblems
at $\tau$. Indeed, this happens either when we count $h$ in some node $u$ of $T_\tau$
that contains $h^*$ and is missed by $q^*$ (and then again $q\in h$ for sure),
or else we count $h$ in the $c^*$ counter of the leaf $u$, at the bottom-level
$k^*$ of $T_\tau$, whose dual box contains $h^*$. In this case $q^*$ crosses this
box. Assuming, as above, that $\tau_v = [0,\delta_1]^d$, the same argument given in
the proof of part (b) implies that the vertical distance from $h^*$ to $q^*$ is
smaller than $\eps$, and therefore so is the real distance from $q$ to $h$, as claimed.
$\Box$

\subparagraph*{Preprocessing and storage.}
Suppressing the expansion of the primal octree at nodes that are not crossed by any
boundary hyperplane makes the storage that it requires $O\left(\frac{n}{\delta_1^{d-1}}\right)$,
and it can be constructed in $O\left(\frac{n}{\delta_1^{d-1}}\right)$ time.
Fix a primal bottom-level leaf cube $\tau=\tau_v$, and put $n_\tau := |S_\tau|$.
It takes $O\left(n_\tau \log \frac{1}{\delta_2}\right)$ time and space to construct $T_\tau$.
(Similar to the primal setup, we prune $T_\tau$ so as not to explicitly represent nodes whose
boxes do not contain any dual point. Note that the constant of proportionality here, as well
as in subsequent bounds, depends exponentially on $d$.)
Since we have $\sum_\tau n_\tau = O(n/\delta_1^{d-1})$,
over all $k$-level leaf cubes $\tau$ of the primal tree $T$, we get that the total construction time
of all dual structures is $O\left(\frac{n}{\delta_1^{d-1}} \log \frac{1}{\delta_2}\right)$,
and this also bounds their overall storage.

When we answer a query $q$, it takes $O\left(\log \frac{1}{\delta_1}\right)$ time to
find the leaf $v$ in $T$ whose cube $\tau_v$ contains $q$ and add up the counters of
the nodes encountered along the path, and then, assuming $v$ to be a bottom-level leaf,
$O\left(\frac{1}{\delta_2^{d-1}}\right)$ time to trace $q^*$ in
$T_{\tau_v}$ and add up the appropriate counters. The total cost of a query is thus
\[
O\left( \frac{1}{\delta_2^{d-1}} + \log \frac{1}{\delta_1} \right) ,
\]
and the total time for $m$ queries is
${\displaystyle O\left(m \left(\frac{1}{\delta_2^{d-1}} + \log \frac{1}{\delta_1}\right)\right)}$.
It is easy to see that the term $\log\frac{1}{\delta_1}$ dominates only when $\delta_2$
is very close to $1$. Specifically this happens when
${\displaystyle \delta_2 = \Omega\left( \frac{1}{\left(\log\frac{1}{\eps}\right)^{1/(d-1)}}\right)}$.

\subparagraph*{Analysis.}
Let $m$ denote the number of queries that we want (or expect) to handle.
The values of $\delta_1$ and $\delta_2$ that nearly balance the construction time with the
total time for $m$ queries, under the constraint that $\delta_1\delta_2=\eps$, are
(ignoring the issue of possible dominance of the term $\log\frac{1}{\delta_1}$
in the query cost)
\[
\delta_1 = \left(\frac{n}{m}\right)^{\frac{1}{2(d-1)}} \sqrt{\eps} ,\quad\quad
\delta_2 = \left(\frac{m}{n}\right)^{\frac{1}{2(d-1)}} \sqrt{\eps} ,
\]
% \[
% \delta_1 = \left(\frac{n}{m}\right)^{\frac{1}{2(d-1)}} \sqrt{\eps} \log^{\frac{1}{2(d-1)}} \frac{n}{\eps^{d-1} m} ,\quad\quad
% \delta_2 = \left(\frac{m}{n}\right)^{\frac{1}{2(d-1)}} \sqrt{\eps} \frac{1}{\log^{\frac{1}{2(d-1)}} \frac{n}{\eps^{d-1} m}} ,
% \]
and the cost is then
\begin{equation} \label{eq:bd1d}
\tilde{O}\left( \frac{\sqrt{mn}}{\eps^{(d-1)/2}} \right) .
\end{equation}
% \begin{equation} \label{eq:bd1d}
% O\left( \frac{\sqrt{mn}}{\eps^{(d-1)/2}} \log^{1/2}\frac{n}{\eps^{d-1} m} \right) .
% \end{equation}
For this to make sense, we must have $\eps \le \delta_1,\;\delta_2 \le 1$,
which holds when
\[
n\eps^{d-1} \le m \le \frac{n}{\eps^{d-1}} ,
\]
% Equivalently, putting $x=\eps^{d-1} m/n$, we require
% \[
% \eps^{2(d-1)} \log\frac{1}{x} \le x \le \log\frac{1}{x} ,
% \]
which means that
\[
% c_1\eps^{2(d-1)}\log\frac{1}{\eps} \le x \le c_2 ,\qquad\text{or}\qquad
c_1\eps^{d-1} n
% \log\frac{1}{\eps} 
\le m \le \frac{c_2 n}{\eps^{d-1}} ,
\]
for suitable absolute constants $c_1$, $c_2$.
%Note that when $m$ is close to the upper bound of this range,
%$\log\frac{1}{\delta_1}$, which is then $\log\frac{1}{\eps}$,
%dominates $\frac{1}{\delta_2^{d-1}}$, and the overall cost of the queries becomes
%$O(m\log\frac{1}{\eps})$, a term that will appear later in the overall
%bound anyway.

When $m < c_1n\eps^{d-1}$, we only work in the dual, for a cost of
\begin{equation} \label{eq:bd2d}
O\left(\frac{m}{\eps^{d-1}}+n\log\frac{1}{\eps}\right) = \tilde{O}\left( n \right) ,
\end{equation}
and when $m > c_2\frac{n}{\eps^{d-1}}$, we only work in the primal space, for a cost of
\begin{equation} \label{eq:bd3d}
O\left(\frac{n}{\eps^{d-1}}+m\log\frac{1}{\eps}\right) =
\tilde{O}\left( m \right) .
\end{equation}
Hence the total cost of $m$ queries, adding up the bounds in
(\ref{eq:bd1d}), (\ref{eq:bd2d}) and (\ref{eq:bd3d}), is
\begin{equation} \label{eq:bdd}
\tilde{O}\left( \frac{\sqrt{mn}}{\eps^{(d-1)/2}} + n + m \right) .
\end{equation}
% \begin{equation} \label{eq:bdd}
% O\left( \frac{\sqrt{mn}}{\eps^{(d-1)/2}} \log^{1/2}\frac{n}{\eps^{d-1} m}
% + n \log\frac{1}{\eps} + m \log\frac{1}{\eps}\right) .
% \end{equation}

%%%%%%%%%%%%%%%%%%%%%%%%%%%%%%%%%%%%%%%%%%%%%%
\subparagraph*{Approximating the maximum depth.}
We can use this data structure to approximate the maximum depth as follows.
For each primal grid cube $\sigma$ of side length $\frac{\eps}{2\sqrt{d}}$,
pick its center $q_\sigma$, compute
$d^-(q_\sigma)$ and $d^+(q_\sigma)$, using our structure,
and report the centers $q^-$ and $q^+$ that achieve $\max_\sigma d^-(q_\sigma)$ and $\max_\sigma d^+(q_\sigma)$, respectively.
%More precisely, for each center point $q_\sigma$ we compute and add up $O(1)$
%sub-counts for each of $d^-(q_\sigma)$, $d^+(q_\sigma)$,
%one for each subset of input halfspaces whose boundary hyperplanes
%have normals in a corresponding restricted cap on $\sph^{d-1}$.
In this application the number of queries is $m = O\left( 1/\eps^d \right)$.

The following lemma asserts lower bounds the
$d^-$ and $d^+$ value of a grid center.

%%%%%%%%%%%%%%%%%%%%%%%%%
\begin{lemma} \label{half-correctd}
(a) Let $q$ be an arbitrary point in $Q$, and let $\sigma$ be the
grid cube of size $\frac{\eps}{2\sqrt{d}}$ that contains $q$. Then we have
\begin{align*}
d_\eps^-(q)  \le d^-(q_\sigma)  \; \text{ and }\;
d_{\eps/2}^-(q)  \le d^+(q_\sigma)  .
\end{align*}

\noindent
(b) In particular, let $\qmx$ be a point of maximum (exact) depth in $\A(S)$,
and let $\sigma$ be the grid cube of size $\frac{\eps}{2\sqrt{d}}$ that contains $\qmx$.
Then we have
\begin{align*}
d_\eps^-(\qmx)  \le d^-(q_\sigma)  \; \text{ and }\;
d_{\eps/2}^-(\qmx)  \le d^+(q_\sigma).
\end{align*}
\end{lemma}
%%%%%%%%%%%%%%%%%%%%%%%%%
\noindent{\bf Proof.}
We only prove (a), since (b) is just a special case of it.
We establish each inequality separately.

\medskip
\noindent{\bf (i)}
$d_\eps^-(q) \le d^-(q_\sigma)$:
Let $h$ be a halfspace that contains $q$, so that $q$ lies at distance
greater than $\eps$ from $\bd h$. Assume without loss of generality that
the inward unit normal of $\bd h$ is in the cap $C_\varphi$ around
the `north pole' of $\sph^{d-1}$.
Since the distance between $q$ and $q_\sigma$ is
at most $\eps/4$, $q_\sigma$ also lies in $h$.

If we do not count $h$ in $d^-(q_\sigma)$
during the primal and dual then $q_\sigma^*$ crosses the bottom-level (dual) box that contains $h^*$.
As in the proof of Lemma~\ref{hs-correctd}(b,c), this implies that the
vertical distance between $q_\sigma$ and $\bd h$ is at most $\frac34\eps$.
Since  the distance between $q$ and $q_\sigma$ is at most $\frac{\eps}{4}$,
so it follows, using the triangle inequality, that the distance from $q$ to
$\bd h$ is at most $\eps$, a contradiction that establishes the claim.

\medskip
\noindent{\bf (ii)}
$d_{\eps/2}^-(q) \le d^+(q_\sigma)$:
Let $h$ be a halfspace that contains $q$ and $q$ lies at distance
at least $\eps/2$ from $\bd h$. In this case it is clear that $h$ contains
$q_\sigma$, so $h$ will be counted in $d^+(q_\sigma)$ by the preceding arguments.
No assumption on the grid size is needed here.
$\Box$

\subparagraph*{In summary,}
using Lemma~\ref{half-correctd}, the analysis of the preprocessing-and-query procedure
(culminating in the bound in (\ref{eq:bdd})), and the fact that here we have
$m = O\left(1/\eps^d\right)$, we obtain the following summary results of this section.

%%%%%%%%%%%%%%%%%%%%%%%%%%
\begin{theorem} \label{th:depthd}
Let $S$ be a set of $n$ halfspaces in $\reals^d$ and let $\eps>0$ be an error parameter.
We can construct a data structure such that, for a query point $q$ in the unit cube
$[0,1]^d$, we can compute two numbers $d^-(q)$, $d^+(q)$ that satisfy
\[
d_\eps^-(q) \le d^-(q) \le d(q) \le d^+(q) \le d_\eps^+(q) .
\]
Denoting by $m$ the number of queries that we expect the structure to perform,
we can construct the structure so that its preprocessing cost and storage, and the time
it takes to answer $m$ queries, are all
\[
\tilde{O}\left( \frac{\sqrt{mn}}{\eps^{(d-1)/2}} + n + m \right) .
\]
\end{theorem}

%%%%%%%%%%%%%%%%%%%%%%%%%%
\begin{theorem} \label{th:maxdepthd}
Let $S$ be a set of $n$ halfspaces in $\reals^d$ and let $\eps>0$ be an error parameter. We
can compute grid centers $q^-$ and $q^+$ such that if $\qmx$ is a point at maximum depth then
\begin{align*}
d_\eps^-(\qmx)  \le d^-(q^-) \;\text{ and }
d_{\eps/2}^-(\qmx)  \le d^+(q^+)  .
\end{align*}
The running time is
\[
\tilde{O}\left( \frac{\sqrt{n}}{\eps^{d-1/2}} n) + n + \frac{1}{\eps^d} \right) .
\]
% \[
% O\left( \frac{\sqrt{n}}{\eps^{d-1/2}} \log^{1/2}(\eps n)
% + n \log\frac{1}{\eps} + \frac{1}{\eps^d} \log\frac{1}{\eps}\right) .
% \]
\end{theorem}

\noindent{\bf Remarks.}
{\bf (i)} The bound in Theorem~\ref{th:depthd} is better than
the naive bound ${\displaystyle O\left(\frac{n}{\eps^{d-1}} + m\log\frac{1}{\eps} \right)}$,
obtained when using the primal-only approach, when ${\displaystyle m = \tilde{O}\left( \frac{n}{\eps^{d-1}} \right)}$.
The bound in Theorem~\ref{th:maxdepthd} is better than the naive bound
${\displaystyle O\left(\frac{n}{\eps^{d-1}} + \frac{1}{\eps^d}\log\frac{1}{\eps} \right)}$,
obtained when using the primal-only approach, when ${\displaystyle n = \tilde{\Omega}\left( \frac{1}{\eps} \right)}$.

\medskip
\noindent {\bf (ii)}
As already discussed, a priori, in both Theorems \ref{th:depthd} and \ref{th:maxdepthd},
the output counts $d^-$, $d^+$ (or $d^-(q^-)$, $d^+(q^+)$) could vary significantly
from the actual depth $d(q)$ (or maximum depth). Nevertheless, such a discrepancy is caused
only because the query point (or the point of maximum depth) lies too close to the
boundaries (either inside or outside) of many halfspaces in $S$.

%%%%%%%%%%%%%%%%%%%%%%%%%%%%%%%%%%%%
\section{Approximate depth for simplices in higher dimensions} \label{app:apxsimp}

The results of Section~\ref{sec:apxtri} (and Appendix~\ref{app:apxtri}) can be extended to higher dimensions.
To simplify the presentation, we describe the case
of three dimensions in detail, and then comment on the extension to any higher dimension.

\subparagraph*{Simplices in three dimensions.}
Our input consists of $n$ simplices in the unit cube $Q=[0,1]^3$.
%\footnote{%
%  With some extra care, we can also handle the case where the simplices only partially
%  overlap $Q$.}
Let $\sigma$ be an input simplex. We represent $\sigma$ as a signed union
involving $O(1)$ regions, so that $\sigma$ is the disjoint union of some
of these regions minus the disjoint union of the others, and so that each
of these regions has at most two faces that are not axis-parallel.
To describe the decomposition, assume for the moment that the coordinate
frame is fixed. We note that, by assumption, all the input simplices lie fully
above the $xy$-plane.
We consider each facet $f$ of $\sigma$ and project it onto the $xy$-plane,
denoting the projection as $f'$. Apply to $f'$ the planar representation of
Section~\ref{sec:apxtri}, writing it as the signed union of three vertical
trapezoids, so that two of them are positive and one is negative, or the other
way around, and their signed union is such that the positive trapezoids
participate in the union and the negative ones are subtracted from it.
Now we lift each of these trapezoids $\tau$ to a $z$-vertical prism whose
floor is $\tau$ and whose ceiling is contained in the plane supporting $f$.
(In general, the ceiling only overlaps $f$, and may even be disjoint from $f$
if $\tau$ is a negative trapezoid.) If $f$ belongs to the upper boundary
of $\sigma$, each prism inherits the sign of its base trapezoid, and if
$f$ belongs to the lower boundary of $\sigma$, each prism gets the opposite
sign of that of its base trapezoid. We note that each prism has the promised
shape: It has (at most) two facets that are not fully axis-parallel: one is
its ceiling, and the other is the lifting of the slanted edge of its base.
In general, the ceiling is not parallel to any coordinate direction, whereas
the second facet is parallel (only) to the $z$-axis.

One can show that $\sigma$ is the signed union of all the resulting prisms.
Actually, the following stronger property holds: For a query point $q$, the sum
of the signs of the prisms that contain $q$ (of the above signed union
of $\sigma$) is $1$ if $q\in\sigma$ and $0$ otherwise.
% \micha{Some proof should be provided?}

The preceding description was for a fixed coordinate frame. In actuality, we face
the same issue as in the case of triangles (Section~\ref{sec:apxtri}), which is
a refinement of a similar issue arising for planes or hyperplanes
(Sections~\ref{sec:apxhalf}, \ref{sec:apxhalfs}). That is,
we want to avoid situations in which (i) the angles between the facets of $\sigma$
and the $z$-direction are too small, or (ii) the angles between the slanted vertical
facets of the prisms and the $y$-direction are too small. In either of these `bad'
situations, we might count in $d_\eps^+(q)$ simplices $\sigma$ for which $q$ lies
outside $\sigma$, at distance much larger than $\eps$ (recall Figure~\ref{offset}).
Extending the arguments in Section~\ref{sec:apxtri}, we can find a positive constant
angle $\beta$ (albeit smaller than the one obtained in the planar case), so that one can
construct $O(1)$ directions on $\sph^2$ and $O(1)$ directions on $\sph^1$, so that we
can assign to each $\sigma\in S$ a pair $(u^{(3)},u^{(2)}) \in \sph^2\times\sph^1$
of directions, so that neither (i) nor (ii) occurs for any prism in the decomposition
of $\sigma$. We construct a separate data structure for each such pair, on the
simplices assigned to that pair, and search all of these structures with the query point.
In what follows we describe the structure for a fixed such pair, or, equivalently,
for a fixed coordinate frame.

Let $P$ (resp., $N$) denote the collection of all prisms with a
positive (resp., negative) sign. We fix one of these collections, say $P$,
and construct the following data structure for approximate depth queries
with respect to the prisms in $P$.

We first construct a segment tree on the $x$-projections of the prisms of $P$.
When we query with a point $q$, we search the tree with its $x$-projection $q_0$,
visiting $O(\log n)$ nodes. The collection of the prisms that are stored at these
nodes coincides with the collection of all prisms $\tau\in P$ such that
the $x$-span of $\tau$ contains $q_0$. Note that each prism arises in at most
one node of the search path.

We now construct the following data structure for each node $v$ of the
segment tree, on the corresponding set $P_v$ of prisms stored at $v$.
As in the previous sections, the structure has a primal part and a dual part.
The primal part is an octree constructed on the planes supporting both slanted
facets of each of the prisms of $P_v$, in a similar manner to the case of (hyper)planes.
Each node maintains a
counter that stores the number of prisms that fully contain its associated
cube but do not fully contain the cube of its parent. Nodes that are not crossed
by the boundary of any prism become shallow leaves and are not expanded further
(nor do they have a dual counterpart). The primal tree is constructed up to a
depth where the cube of each leaf is of side length $\delta_1$. The cost of
constructing the primal octree is $O(n/\delta_1^2)$.

At each deep leaf $v$ of this octree, we pass to a dual substructure, which
has two levels, each storing one of the two slanted facets of each of the
prisms associated with $v$ (namely, prisms with at least one slanted
facet crossing $\tau_v$), which is mapped to a dual point.
However, the two dual points live in different dimensions: the ceiling is
mapped to a point in $\reals^3$, whereas the other slanted $z$-vertical facet is
mapped to a point in the plane (as its equation is independent of $z$).
In the presentation that follows we assume that both slanted facets of the
prism cross $\tau_v$; the cases where only one of them crosses $\tau_v$
are easier to handle---in such cases one needs only one level of the dual
structure.

The first dual level handles, say, the ceilings of the prisms as
points in $\reals^3$. It follows the dual structure for the case of planes,
described in Section~\ref{sec:apxhalf}, except that each node $v$ of
the structure, instead of storing a counter, collects all the relevant
halfspaces, moves to the set of the corresponding slanted vertical facets of
the same prisms, and processes this set into a substructure associated
with $v$ at the second dual level. However, each of the deep leaves of
the first dual level still stores a counter of the number of prisms for
which the point dual to the ceiling of the prism lies in the region of
the leaf, and these prisms are not passed to the second dual level.

The second dual level handles the slanted vertical faces of the prisms as points
in the plane. Here we follow the structure of Section~\ref{sec:apxhalfs} verbatim,
storing a counter at each node, as described there.

\medskip
\noindent{\bf Remark.}
Note that the segment tree is in fact a refined and improved version of what
otherwise would be a third dual stage of the construction (on the
$x$-projections of the prisms). It allows us to control in an exact
manner the relation between the $x$-coordinate of the query point
and the $x$-spans of the simplices, leaving us with handling of the
$\eps$-deviations only in the $y$- and $z$-directions.

To answer a query with a point $q$, we first query the segment tree
with the $x$-coordinate $q_0$ of $q$, to retrieve the $O(\log n)$
nodes that $q_0$ reaches. The prisms stored at these nodes
are precisely those for which $q$ lies in the correct side of each
of the axis-aligned facets of the prism. (Note that here we obtain
the exact set of these prisms.) It therefore remains to count
the number of those prisms for which $q$ lies on the correct side of
each of their two slanted facets, within the usual inside / outside
deviation error of $\eps$.

To do so, at each node $v$ of the segment tree
that $q_0$ reaches, we first query the primal octree of the structure
associated with $v$, add up the counters that are stored at the nodes
that $q$ reaches, adding that sum to
both $d^-(q)$ and $d^+(q)$, and then pass to the dual structure at the
deep leaf that $q$ reaches, with the set of prisms stored at that leaf.

As we query the first dual level, at each node $v$ that $q^*$ (now a
plane in three dimensions) reaches, we pass to the second dual level,
constructed over the slanted vertical facets of the corresponding prisms,
and query it too with $q^*$ (now a line in the plane). However, at the
deep leaves of the first level, we do not pass to the second level and
just add the counters at these leaves to $d^+(q)$.

At the second dual level, at each node $v$ that $q^*$ reaches,
we add the counter that it stores to both $d^-(q)$ and $d^+(q)$,
except for the counters at the leaves which are only added to $d^+(q)$.

Performing this procedure over all relevant nodes of the segment tree,
and over the $O(1)$ choices of the coordinate frame, we add up the
counters obtained from these substructures, and output the resulting
values $d^-(q)$ and $d^+(q)$ as $\pi^-(q)$ and $\pi^+(q)$, respectively.

We construct a similar data structure for the prisms in $N$, and
query it with the point $q$ exactly as above, obtaining corresponding
overestimate and underestimate for the depth of $q$ in $N$, which
we now denote as $\nu^+(q)$ and $\nu^-(q)$, respectively.

As in Section~\ref{sec:apxtri}, we return the values
\begin{align} \label{dpinud}
d^-(q) & := \pi^-(q) - \nu^+(q) \\
d^+(q) & := \pi^+(q) - \nu^-(q) \nonumber .
\end{align}

\subparagraph*{Analysis.}
Recall that the primal octree is constructed up to cubes of side length
$\delta_1$. Each dual octree, in both levels, is expanded till we reach
a resolution refinement of $\delta_2$, as in the preceding sections,
with $\delta_1\delta_2 = \eps$. The preprocessing cost, summed over
all nodes of the segment tree (and over all coordinate frames), is
\[
O\left( \frac{n}{\delta_1^2} \log^2 \frac{1}{\delta_2} \log n\right) .
\]
The cost of searching in a fixed primal tree (for $m$ queries) is
${\displaystyle O\left( m\log\frac{1}{\delta_1}\right)}$, and the cost of searching
in the dual structures is ${\displaystyle O\left( \frac{m}{\delta_2^3}\right)}$,
because each query is a two-level
query, where, as already said, the first level is with a dual plane $q^*$ that
crosses $O(1/\delta_2^2)$ regions of the first level, and the second level
of the query is with a dual line that crosses $O(1/\delta_2)$ regions of the
second level for each first-level node that $q^*$ reaches. The overall
cost of the structure, on $n$ simplices and $m$ queries, summed over the
nodes of the segment tree and the coordinate frames, is therefore
\[
O\left( \frac{n}{\delta_1^2}\log^2\frac{1}{\delta_2}
+ m\left(\frac{1}{\delta_2^3} + \log\frac{1}{\delta_1} \right) \right) .
\]
Balancing (roughly) the two terms, under the constraint $\delta_1\delta_2 = \eps$, and
assuming that $\frac{1}{\delta_2^3}$ dominates $\log\frac{1}{\delta_1}$, yields
\[
\delta_2 = \left( \frac{m\eps^2}{n} \right)^{1/5} 
% \cdot \frac{1}{\log^{2/5} \left( \frac{n}{m\eps^2}\right)}
\qquad\text{and}\qquad
\delta_1 = \left( \frac{m\eps^3}{n} \right)^{1/5} ,
%  \log^{2/5} \left( \frac{n}{m\eps^2}\right) ,
\]
making the overall performance of the structure
\[
\tilde{O}\left( \frac{m^{2/5}n^{3/5}}{\eps^{6/5}} \right) .
\]
% \[
% O\left( \frac{m^{2/5}n^{3/5}}{\eps^{6/5}} \log^{6/5} \left( \frac{n}{m\eps^2} \right) \log n \right) .
% \]
As in the preceding sections, this holds provided that $\eps \le \delta_1,\;\delta_2 \le 1$,
which holds when
\[
c_1 n\eps^3 \le m \le \frac{c_2 n}{\eps^2} ,
\]
% \[
% c_1 n\eps^3 \log^{2/5}\frac{1}{\eps} \le m \le \frac{c_2 n}{\eps^2} ,
% \]
for suitable constants $c_1$, $c_2$. One can show that when $m$ is larger we get the bound
$\tilde{O}\left(m\right)$, and when $m$ is smaller we get the bound
$\tilde{O}\left(n\right)$. Altogether, we obtain the bound
\[
\tilde{O}\left( \frac{m^{2/5}n^{3/5}}{\eps^{6/5}} + m + n \right) .
\]
% \[
% O\left( \left(
% \frac{m^{2/5}n^{3/5}}{\eps^{6/5}} \log^{6/5} \left( \frac{n}{m\eps^2} \right) +
% m\log\frac{1}{\eps} + n\log^2\frac{1}{\eps} \right) \log n \right) .
% \]

A suitably adapted version of the analysis in the preceding sections shows that
\begin{align*}
\pi_\eps^-(q) & \le \pi^-(q) \le \pi(q) \le \pi^+(q) \le \pi_\eps^+(q) \\
\nu_\eps^-(q) & \le \nu^-(q) \le \nu(q) \le \nu^+(q) \le \nu_\eps^+(q) ,
\end{align*}
where (i) $\pi(q)$ is the depth of $q$ in $P$,
(ii) $\pi_\eps^-(q)$ is the number of all prisms $\tau\in P$ such
that $q \in \tau$ and the distance from $q$ to the slanted
part of the boundary of $\tau$ is at least $\eps$, and
(iii) $\pi_\eps^+(q)$ is the number of all prisms $\tau\in P$ such
that (iii.a) $q$ lies in the offset prism $\tau'$ of $\tau$ obtained by shifting
the two planes supporting the slanted facets of $\tau$ by distance $\eps$ away
from $\tau$, and (iii.b) $q$ lies in the $x$-span of $\tau$.
$\nu(q)$, $\nu_\eps^-(q)$, $\nu_\eps^+(q)$ are defined analogously for $N$.

\begin{figure}[htb]
\begin{center}
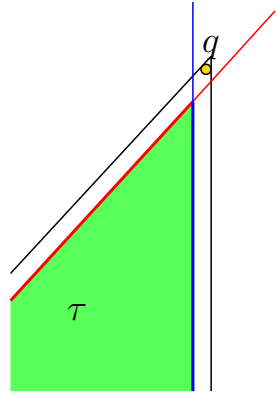
\caption{\sf $q$ lies outside $\tau$, the distance from $q$ to each of
the planes supporting the two slanted facets of $\tau$ is at most $\eps$,
but its distance from $\tau$ is larger than $\eps$.}
\label{fig:outside}
\end{center}
\end{figure}

Here too, $\pi_\eps^+(q)$ and $\nu_\eps^+(q)$ are defined slightly differently
from the way they are defined for hyperplanes---this is the same issue, already
mentioned, that arose in the case of triangles in the plane (see Section~\ref{sec:apxtri}).
That is, when $q$ is outside $\tau$, the fact that the distance from $q$
to each of the two planes supporting the slanted facets of $\tau$ is at most $\eps$
does not necessarily guarantee that its distance from $\tau$ is at most $\eps$;
see Figure~\ref{fig:outside} (and recall also Figure~\ref{offset}).
Nevertheless, the choice of $O(1)$ canonical
coordinate frames and the assignment of simplices to frames allows us to
ensure that the distance is at most some fixed multiple of $\eps$.

% \micha{Not sure we need this figure?}

In contrast, for $\pi_\eps^-(q)$ and $\nu_\eps^-(q)$, being at distance
at least $\eps$ from each of the two planes supporting the slanted facets
of $\tau$ is equivalent to being at distance at least $\eps$ from $\bd\tau$.

As in Section~\ref{sec:apxtri}, the values $d^-(q)$, $d^+(q)$ that we return,
as in (\ref{dpinud}), satisfy
\[
\pi_\eps^-(q) - \nu_\eps^+(q)
\le d^-(q) \le \pi(q) - \nu(q) \le d^+(q) \le \pi_\eps^+(q) - \nu_\eps^-(q) .
\]
It follows, by construction, that $\pi(q) - \nu(q)$ is the real depth $d(q)$
of $q$ in $S$.
Similarly, $\pi_\eps^-(q) - \nu_\eps^+(q)$ counts all simplices $\sigma$
for which (i) $q$ lies in a unique prism $\tau^+$ of $P$ that participates
in the signed union decomposition of $\sigma$, at distance at least $\eps$
from the slanted portion of its boundary, and (ii) for any prism
$\tau^-$ of $N$, $q$ lies outside $\tau^-$, at distance larger than $\eps$
from the slanted portion of its boundary.
Hence $\pi_\eps^-(q) - \nu_\eps^+(q)$ counts all simplices $\sigma$
that (i) contain $q$, (ii) $q$ lies at distance at least $\eps$ from $\bd\sigma$,
and (iii) $q$ lies in the $x$-span of $\sigma$.

A similar argument shows that $\pi_\eps^+(q) - \nu_\eps^-(q)$
counts all simplices $\sigma$ such that their $\eps$-offset contains $q$
and $q$ lies in the $x$-span of the simplex.

In summary, we obtain the first main results of this section,
stated as Theorems~\ref{simp3d} and \ref{simp3dmx} in the main part of the paper.
The second result is obtained from the first, arguing as in the preceding sections.

%%%%%%%%%%%%%%%%%%%%%%%%%%%%%%
\subparagraph*{Higher dimensions.}
We only sketch the extension to higher dimensions. Let $S$ be a set of
$n$ simplices in the unit cube in $\reals^d$ (now for $d\ge 4$).
Extending recursively the decomposition scheme in two and three dimensions,
we represent each simplex $\sigma$ in $S$ as the signed union of prisms, where
each prism has at most $d-1$ slanted facets, where the first facet is aligned
with an original facet of $\sigma$, the second facet is parallel to the $x_d$-axis,
the third is parallel to the $x_{d-1}x_d$-plane, and so on. Thus when we dualize
these facets, we end up with a sequence of $d-1$ points, where the $j$-th point
lies in $\reals^{d-j}$, for $j=0,\ldots,d-2$.

As in the three-dimensional case, we want to make sure that none of the
slanted facets of a prism is too steep, and we enforce it by creating $O(1)$
coordinate frames, assign each simplex of $S$ to a suitable frame, and repeat
both preprocessing and queries for each frame (and the simplices assigned to it).

We construct a segment tree on the $x_1$-spans of the prisms; this `gets rid' of
the two $x_1$-orthogonal facets of each prism. At each node of the tree we
construct a data structure consisting of one primal level (in dimension $d$),
on all the slanted facets of each prism, and of $d-1$ dual levels, in dimensions
$2,\ldots,d$, catering to the different dual points of the slanted facets of
each prism.\footnote{%
  As in the three-dimensional case, the segment tree can be regarded as an
  additional one-dimensional level of the structure.}
Queries are performed in full analogy to the three-dimensional case.
The overall cost of the structure is
\begin{align*}
& O\left( \left( \frac{n}{\delta_1^{d-1}} \log^{d-1} \frac{1}{\delta_2} +
m\left(\frac{1}{\delta_2^{(d-1)+(d-2)+\cdots +1}} + \log\frac{1}{\delta_1} \right) \right) \log n\right) \\
& = O\left( \left( \frac{n}{\delta_1^{d-1}} \log^{d-1} \frac{1}{\delta_2} +
m\left(\frac{1}{\delta_2^{d(d-1)/2}} + \log\frac{1}{\delta_1} \right) \right) \log n\right) .
\end{align*}
% \begin{align*}
% & O\left( \left( \frac{n}{\delta_1^{d-1}} \log^{d-1} \frac{1}{\delta_2} +
% m\left(\frac{1}{\delta_2^{(d-1)+(d-2)+\cdots +1}} + \log\frac{1}{\delta_1} \right) \right) \log n\right) \\
% & = O\left( \left( \frac{n}{\delta_1^{d-1}} \log^{d-1} \frac{1}{\delta_2} +
% m\left(\frac{1}{\delta_2^{d(d-1)/2}} + \log\frac{1}{\delta_1} \right) \right) \log n\right) .
% \end{align*}
Balancing (roughly) the two terms, under the constraint $\delta_1\delta_2 = \eps$, ignoring
the case where $\log\frac{1}{\delta_1}$ dominates the coefficient of $m$, yields
\begin{align*}
\delta_2 & = \left( \frac{m\eps^{d-1}}{n} \right)^{2/(d+2)(d-1)}
% \cdot \frac{1}{\log^{2/(d+2)}\left(\frac{n}{m\eps^{d-1}}\right) }
\qquad\text{and} \\
\delta_1 & = \left( \frac{n\eps^{d(d-1)/2}}{m} \right)^{2/(d+2)(d-1)} ,
% \log^{2/(d+2)}\left(\frac{n}{m\eps^{d-1}}\right) ,
\end{align*}
making the overall performance of the structure
\[
\tilde{O}\left( \frac{m^{2/(d+2)}n^{d/(d+2)}}{\eps^{d(d-1)/(d+2)}} \right) .
\]
% \[
% O\left( \frac{m^{2/(d+2)}n^{d/(d+2)}}{\eps^{d(d-1)/(d+2)}}
% \log^{\frac{d(d-1)}{d+2}} \left( \frac{n}{m\eps^{d-1}}\right) \log n \right) .
% \]
Again, this holds as long as $m$ is not too small nor too large. Handling these
extreme cases too, we obtain the perfomance bound, for $n$ simplices and $m$ queries,
\begin{equation} \label{eq:simpdd}
\tilde{O}\left( \frac{m^{2/(d+2)}n^{d/(d+2)}}{\eps^{d(d-1)/(d+2)}} + m + n \right) .
\end{equation}
% \begin{equation} \label{eq:simpdd}
% O\left( \left( \frac{m^{2/(d+2)}n^{d/(d+2)}}{\eps^{d(d-1)/(d+2)}}
% \log^{\frac{d(d-1)}{d+2}} \left( \frac{n}{m\eps^{d-1}}\right)
% + m \log\frac{1}{\eps} + n \log^{d-1}\frac{1}{\eps} \right)\log n \right) .
% \end{equation}
As in the previous sections, one can show that the algorithm is faster than
the earlier approach of \cite{AM00} when $m < \frac{n}{\eps^{d-1}}$.
%\micha{Check the polylog factor, and compare with Fonseca too.}

\subparagraph*{Finding an approximate maximum depth} is done as in the preceding algorithms.
The running time, with $m = 1/\eps^d$, is
\begin{equation} \label{eq:simpddmx}
\tilde{O}\left( \frac{n^{d/(d+2)}}{\eps^{d(d+1)/(d+2)}} + \frac{1}{\eps^d} + n \right) .
\end{equation}
% \begin{equation} \label{eq:simpddmx}
% O\left( \left( \frac{n^{d/(d+2)}}{\eps^{d(d+1)/(d+2)}}
% \log^{\frac{d(d-1)}{d+2}} \left( \eps n\right)
% + \frac{1}{\eps^d} \log\frac{1}{\eps} + n \log^{d-1}\frac{1}{\eps} \right)\log n \right) .
% \end{equation}

%\medskip
%\noindent{\bf Remark.}
%\micha{Comment that an alternative multi-level approach, where the dual
%treats each facet of the original simplex in a separate level
%(and the primal treats all of them simultaneously), gives a worse performance,
%but probably still better than the naive solution. Here too we have the problem
%that when the query is outside the simplex, we test whether its distance from each
%of the hyperplanes supporting a facet is at most $\eps$, but this does not
%guarantee that the distance from the simplex itself is at most $\eps$.}

\section{Implementation}

\begin{figure}[htb]
\begin{center}
\includegraphics[width=120mm]{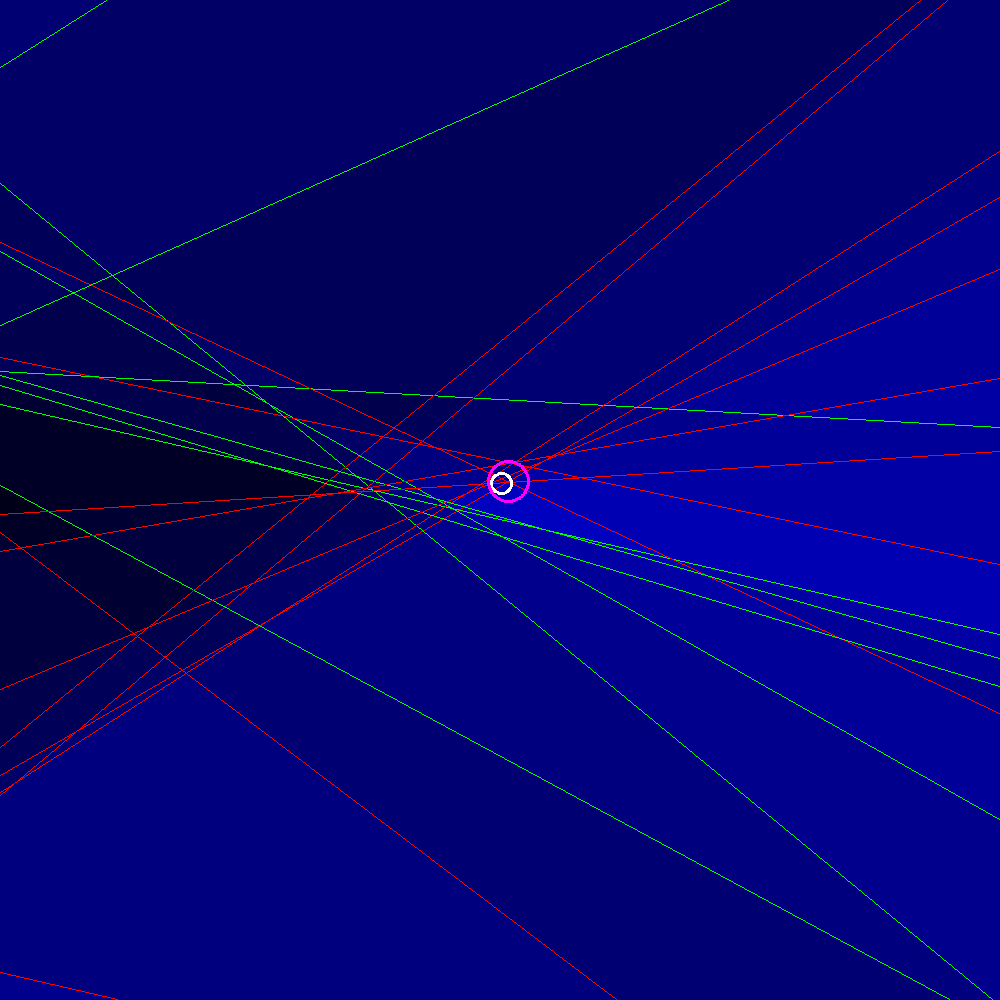}
\caption{Example of the depth (and maximum depth) for $\eps=0.001$ and 20 halfplanes (red are down, green are up). The intensity of the color is proportional to the depth. The maximum depth of the naive algorithm is in white, the maximum depth of the primal dual is in pink. The maximum depth here is 15.}
\label{fig:example}
\end{center}
\end{figure}

\begin{figure}[htb]
\begin{center}
\includegraphics{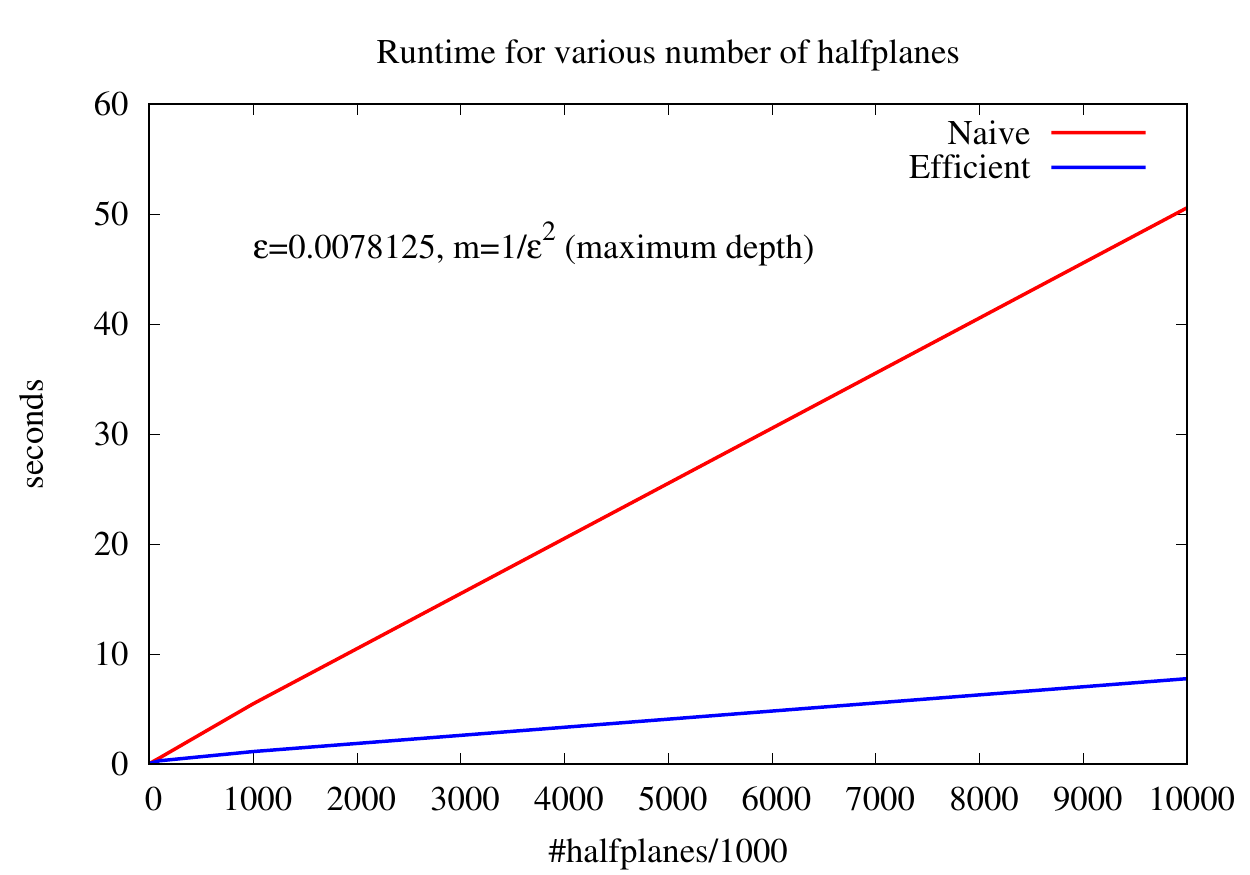}
\caption{Maximum depth for fixed $\eps$ and various $n$}
\label{fig:varn}
\end{center}
\end{figure}

\begin{figure}[htb]
\begin{center}
\includegraphics{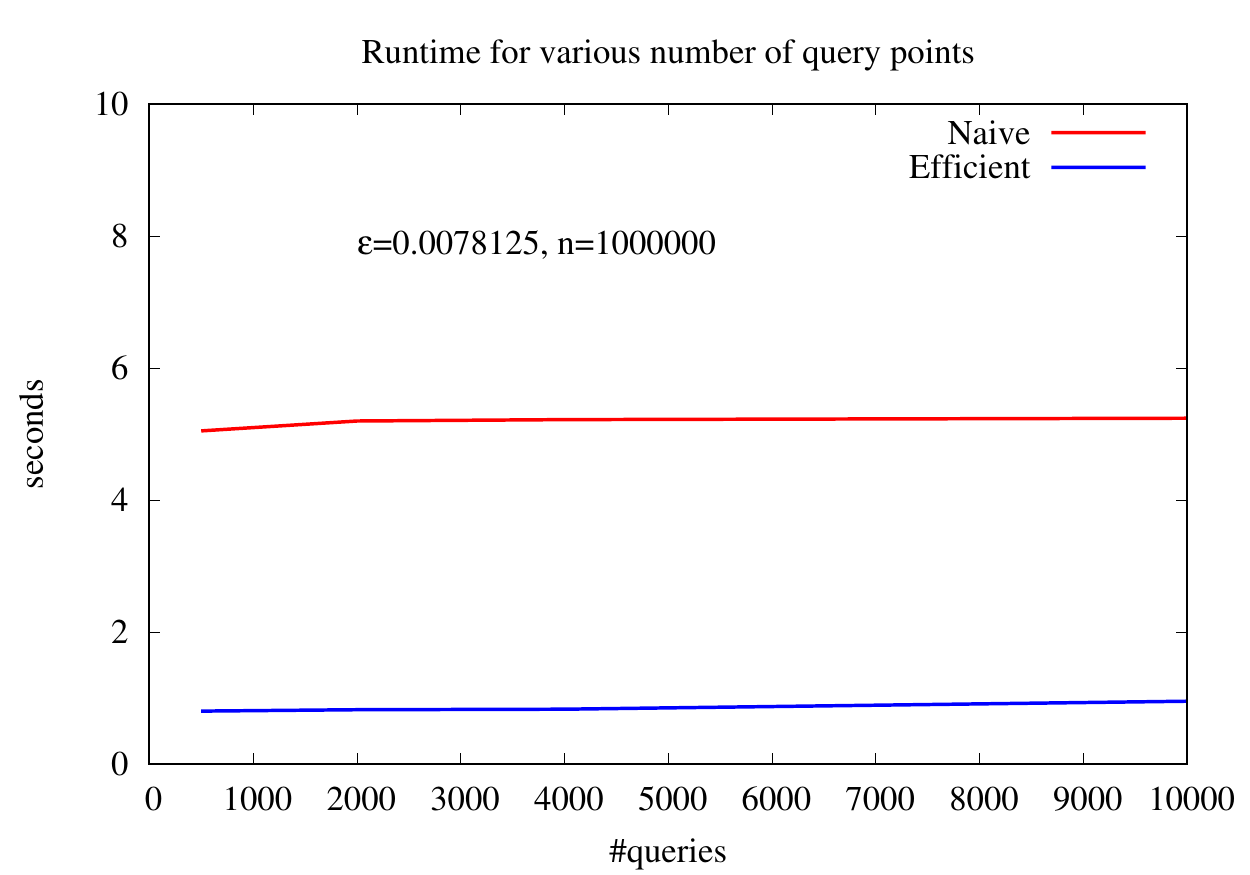}
\caption{The number of queries is vary for fixed $\eps$ and $n$}
\label{fig:varm}
\end{center}
\end{figure}

\begin{figure}[htb]
\begin{center}
\includegraphics{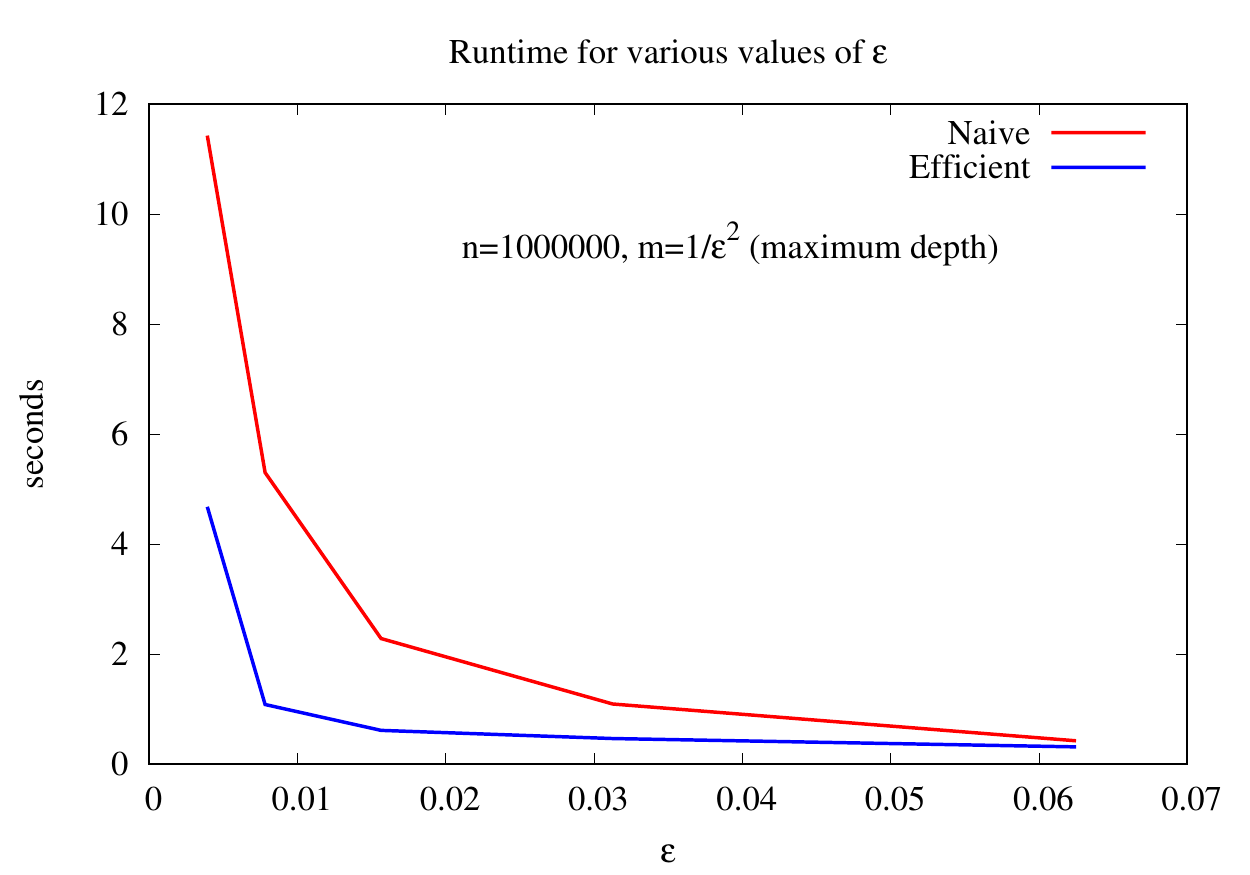}
\caption{Maximum depth for fixed $n$ and various $\eps$ values}
\label{fig:vare}
\end{center}
\end{figure}

We implemented the naive quadtree and the primal-dual algorithm for halfplanes in C++ and evaluated the performance for various parameters. In all tests, $\delta_1$ and $\delta_2$ were automatically selected to the optimal values (depending on the number of halfplanes, $n$, the number of queries, $m$ and $\eps$, see Section~\ref{sec:apxhalf}) and were multiplied by a constant (fixed for all tests) that optimizes the runtime (implementation dependent). In order to make the input better representing real world problems, we created a setup that has a significant maximum depth. $2/3$ of the halfplanes are passing close to the center with uniform random slope in $[-1,1]$ and uniform vertical small shift in $[-0.04,0.04]$. These halfplanes create the significant peak in depth. The other $1/3$ of the halfplanes are uniformly random with slope in [-1,1] and they are crossing $x=0$ at random value in $[0,1]$. These halfplanes are outliers (noise). In Figure \ref{fig:varn}, the runtime for maximum depth (the number of queries is $m=1/\eps^2$) for fixed $\eps$ and increased number of halfplanes is shown. In Figure \ref{fig:varm} we keep the number of halfplanes fixed and increase the number of queries (this evaluation does not apply maximum depth). In Figure \ref{fig:vare}, we keep the number of halfplanes fixed and again apply maximum depth (meaning that the number of queries changes with $\eps$) for various $\eps$ values. Figure \ref{fig:example} is an example of the structure and results from both naive and primal dual maximum depth for 20 halfplanes.

\end{document}